\documentclass[dvipdfm,brazil,12pt]{article} 
\usepackage[ansinew]{inputenc}
\usepackage{amssymb,amsthm,amsfonts,amsmath,natbib}
\usepackage[pdftex]{graphicx}   
\usepackage{epsfig} 
\usepackage{multicol}
\usepackage{multirow}   
\usepackage{scalefnt} 
\usepackage{epstopdf}	
\usepackage{rotating} 
\usepackage{float}   
\usepackage{subfigure}
\usepackage{bbm}
\usepackage{booktabs}

\usepackage{wrapfig}
\usepackage{appendix}
\usepackage{fancyhdr}
\usepackage{url}
\graphicspath{{./Figuras/}} 

\usepackage[paperwidth=215.9mm,paperheight=279.4mm,hmargin={25mm,25mm},vmargin={25mm,25mm}]{geometry}

\renewcommand{\arraystretch}{3.0}



\begin{document}


\begin{center}
	{\Large\bf Exact Bayesian Inference for Geostatistical Models under Preferential Sampling}
\end{center}
\centerline{\bf Douglas Mateus da Silva\footnotemark[1], Dani Gamerman\footnotemark[2]}\centerline\\
\vskip 0.5cm
\footnotetext[1]{Universidade Federal de Minas Gerais,\\ \text{\hspace{7mm}}Av. Antônio Carlos 6627, Pampulha, Belo Horizonte, MG 31270-010, Brazil\\
	\text{\hspace{7mm}}E-mail: douglas\_est@yahoo.com.br}
\footnotetext[2]{Universidade Federal do Rio de Janeiro.}

\begin{abstract}
Preferential sampling is a common feature in geostatistics and occurs when the locations to be sampled are chosen based on information about the phenomena under study. In this case, point pattern models are commonly used as the probability law for the distribution of the locations. However, analytic intractability of the point process likelihood prevents its direct calculation. Many Bayesian (and non-Bayesian) approaches in non-parametric model specifications handle this difficulty with approximation-based methods. These approximations involve errors that are difficult to quantify and can lead to biased inference. This paper presents an approach for performing exact Bayesian inference for this setting without the need for model approximation. A qualitatively minor change on the traditional model is proposed to circumvent  the likelihood intractability. This change enables the use of an augmented model strategy. Recent work on Bayesian inference for point pattern models can be adapted to the geostatistics setting and renders computational tractability for exact inference for the proposed methodology. Estimation of model parameters and prediction of the response at unsampled locations can then be obtained from the joint posterior distribution of the augmented model. Simulated studies showed good quality of the proposed model for estimation and prediction in a variety of preferentiality scenarios. The performance of our approach is illustrated in the analysis of real datasets and compares favourably against approximation-based approaches.
The paper is concluded with comments regarding extensions of and improvements to the proposed methodology.
\end{abstract}

\section{Introduction}\label{sec:intro}
Preferential sampling (PS) occurs when sampled locations are more likely to be chosen than others according to mechanisms that depend on the data generating process. In geostatistics, it is common to ignore the information of the sampling design in the inference step. This is not a problem in a non preferential sampling (NPS) context, where the sampling design is stochastically independent of the spatial process, but the use of the traditional model will provide biased inference and prediction when PS is present.

The NPS model has two main components: a Gaussian process that models the correlated structure of data and a observational specification based on a random observational noise added to the underlying Gaussian process. No information about the process that generated the pattern of the locations (other than their coordinates) is used, since this process is independent of the data process. When PS is present, the stochastic dependence of the point and the Gaussian processes must be both taken into account.

In this way, \cite{diggle2010} proposed a class of models for dealing with PS in geostatistics. A component to explicitly describe
the sampling design was included in the model in addition to the latent Gaussian process and the observational error process of the traditional model. This
component consisted on a Cox process, ie, a spatial Poisson process with intensity based on the same Gaussian process that drives the data.
The authors showed that the addition of this component to the model corrects the resulting inference.
Many papers have been conducted after and based on their study. For example, \cite{pati2011} presented theoretical results of the model in a Bayesian framework; \cite{gelfand2012} analyzed the effect of PS on parameter estimation and prediction, also using a Bayesian approach; \cite{sahddick2014} extended the model to the spatial-temporal context; \cite{ferreiragam} studied the effect of PS on the optimal design; \cite{dinsdale2019} proposed to use Laplace approximation for the likelihood function in a classic approach instead of the Monte Carlo approximation used by \cite{diggle2010}.

The work cited above had to deal with the likelihood function of the inhomogeneous Poisson process. It has no closed form and depends on a integral of the unknown intensity function over the entire continuous domain of the study region.
Approximated methods were considered, based on discretizing the region under study \citep{mollerw}.
Usually in this approximation, a regular grid over the region is used and the real locations of data are approximated by the centroids of the cell that contains the sampled points. Accordingly, the Gaussian process is approximated by a multivariate Gaussian vector over the grid.
Another option for the discretization is via approximation of the integral in the likelihood via quadrature rules
\citep{butler&moffitt1982,inproceedings2017}.

Discretization is a relatively easy solution for the problem but can generate errors that are difficult to quantify and lead to biased estimation \citep{simpson2016}. According to these authors, two main sources of errors are present in the discrete approach: the approximation of the Gaussian process in a discrete domain and the approximation of the real locations by the discrete ones. The last is the dominant source of error in lattice approximation because of the loss of the real information about the locations.

The likelihood function of the Poisson process is the component of the geostatistical model under PS that requires approximated methods. Its intractability impacts the inference and, as consequence, the capability of prediction with the model. Predicting the response variable at unobserved locations is usually the main goal in geostatistics \citep{diggle2007}. Thus, results are also compromised if the path to obtain prediction is compromised. This situation motivates the development of a methodology free from discretization errors.

\cite{geg2018} proposed a methodology for performing exact inference for Cox processes where the intractability of the likelihood function was avoided. This was possible due to an augmentation data approach in which the inhomogeneous Poisson process was embedded as a part of a larger homogeneous Poisson process. As a result, the augmented likelihood function obtained in combination with a bounded link for the intensity leads to a completely tractable model. No approximation was used in their approach, leading to a better and correct inference for Cox processes.

Following this idea, a methodology to perform exact inference in geoestatistics under PS is proposed in this paper, by an adaptation of their results to the geostatistics setting. The proposed method involves the additional restriction of bounded intensity function of the point process. This is the only change required to the traditional PS model in geostatistics. This modification seems not only reasonable in many practical situations but also of minor relevance since most studies are carried out over a bounded region. Therefore, the intensity of interest is bounded to this study region and almost inevitably becomes bounded there even in the potentially unbounded intensity scenario of the traditional approach.

The approach leads to an easily computed likelihood function and requires no approximation of the model, either by discretization or by any other form of approximation. In this way, the methodology provides correct results for inference and prediction. Correct information about the sampled locations is used instead of approximated ones, potentially leading to better estimates of all model parameters. The improvement is also obtained for estimation of the dependence between locations since they depend on the actual sampled points. Also, the methodology works with a continuous spatial domain instead of a discretized one. A study of simulation in PS and NPS contexts will also be presented to compare results from the proposed model and the traditional model and to provide performance assessments.

Thus, the novelty of our contribution lies in: a) proposing a change in the intensity specification and showing how this change allows for
exact model specification in the geostatistics setting, and 2) adapting the inference performed by \cite{geg2018} over point processes for performing inference over the geostatistics setting under PS.
The paper is organized as follows: Section \ref{sec:context} discusses background about the geostatistical model and presents the proposed novel methodology to the PS approach. A simulation study and their main findings are presented in  Section \ref{sec:simul}. Section \ref{sec:application} illustrates the methodology in two real datasets. Finally, Section \ref{sec:discussion} presents our discussion and final remarks.

\section{Context} \label{sec:context}
This section presents a brief review of the geostatistical model under preferential sampling and shows the main difficulty with the existing methodology. Then, the proposed model to make exact inference in the preferential sampling context is presented.

\subsection{Preliminaries}
Let $B$ be a compact region in $\mathbb{R}^m$, $m \geq 1$, the region of study. A point $x$ in $B$ is denoted by $x = (b_1,b_2,...,b_m)$.
The traditional geostatistical model is defined by
\begin{eqnarray}
	Y_i &=& \mu(x_i) + S(x_i) + \epsilon_i, \quad i=1,...n, \label{eq:geo_model}\\
	\mu(x_i) &=& \eta_0 + \eta_1 d_1 (x_i) + ... + \eta_p d_p (x_i), \nonumber
\end{eqnarray}
where $Y_i \equiv Y ( x_i )$ is the observed response at location $x_i$, $\mu(x_i)$ is the deterministic component that represents the expected value of $Y_i$, ($\eta_0, \eta_1,..., \eta_p$) and ($d_1(.), ..., d_p(.)$) are regression coefficients and their respective covariates, $S(.)$ is a stationary Gaussian process with zero mean, constant variance $\sigma^2$  and isotropic correlation function $\rho(x,x')= \rho_\phi (h)$,  where $h = | x - x' |$ is the Euclidean distance between $x$ and $x'$. This process is denoted $GP ( 0 , \sigma^2 , \rho_\phi )$ hereafter.
The observation errors $\epsilon_i$'s are independent of $S$ and usually assumed to be independent Gaussian variables with zero mean and variance $\tau^2$. The model without covariates is a special case of equation (\ref{eq:geo_model}), where $\mu(x) = \eta_0$.

Let $X$ be the spatial point process that models the sampled locations.
In the non-preferential context, $S$ is independent of $X$ (denoted by $[X | S ] = [ X ]$)
and the distribution of $X$ is irrelevant for inference about $S$ and hence for prediction. In the preferential sampling
context, $[X | S ] \neq [ X ]$ and ignoring the dependence between $X$ and $S$ can lead to biased
inference \citep{diggle2010}. In this way, the complete distribution of S, X and Y must be informed.
Usually the process $X|S$ is modeled by a Poisson process with intensity function
\begin{eqnarray}\label{eq:intensity}
	\lambda(x) = g(S(x)),
\end{eqnarray}
where $g$ is a monotonic function. Its likelihood function is given by
\begin{equation}
	f(X|S) \propto  \exp \left\{ -\int_{B}\lambda(\xi)d\xi \right\}  \left[\prod_{i=1}^{n}\lambda(x_i)\right]. \label{eq:poisson_process_likel}
\end{equation}
The main problem of dealing with the Poisson process is that equation (\ref{eq:poisson_process_likel}) is intractable for intensity (\ref{eq:intensity}) when $S$ is unknown. Approximations are required to evaluate it. Such approximations are many times obtained by discretizing the region under study in a grid, in which the intensity is assumed to be constant within each grid cell
\citep{mollerw}. \cite{diggle2010} used a log-Gaussian Cox Process (LGCP) to model $X$, with the link function in equation (\ref{eq:intensity}) given by $g ( u ) = \exp\{\alpha + \beta u \}$. This choice also leads to an intractable likelihood function.

Hereafter, methods with no model approximations are referred to as exact and the exact approach to solve the likelihood intractability is proposed in the next section.

\subsection{Model specification}

The model in this paper retains the basic geostatistical model (\ref{eq:geo_model}) for the
measurements $Y$ and also makes use of a Poisson process with intensity depending on the
underlying process $S$ to describe the preferentiality of the locations $X$. However, the link for the
relation of the intensity of $X$ with the latent process $S$ is rather a
bounded function. This change is fundamental for achieving the exactness of our specification.

Hence, the following model is assumed
\begin{eqnarray}
	Y|X,S &\sim&  N_n(D \eta + S(x), \tau^2 I_n), \label{eq:mBA:Y}\\
	X|S,\lambda &\sim&  PP(\lambda),\\
	\lambda(x) &=&  \lambda^* F(\beta S(x)/\sigma), \label{eq:mBA:lambda}\\
	S|\sigma^2,\phi &\sim&  GP(0, \sigma^2 , \rho_\phi ),\label{eq:mBA:S} \\
	\theta &\sim& \pi(\theta), \quad \theta = (\lambda^*, \eta, \tau^2, \sigma^2, \phi, \beta), \label{eq:mBA:theta}
\end{eqnarray}
where PP means Poisson process, $I_n$ is the identity matrix of order $n$,
$\rho_\phi$ is the correlation function indexed by parameter $\phi$, $F$ is a monotonic function restricted to the unit interval (e.g. a continuous distribution function) and
$\lambda^* = \text{sup}\{\lambda(x)\}$. The probit link $\Phi$ is used for $F$ in the sequel
but any other bounded function, such as the logistic distribution function, could be used in (\ref{eq:mBA:lambda}). These two choices are similar but the probit function is chosen because it is particularly useful for computational purposes.

The following augmented model approach can be used now due to the boundedness of the intensity.
Let $W$ be a homogeneous Poisson process with intensity $\lambda^*$ and suppose that $X$ is obtained by applying Poisson \textit{thinning} in $W$. Thus, the $W$ process can be split in the $X$ and $\tilde{X}$ process according to the thinning mechanism described below
\begin{eqnarray}
	X &=& \{({x},z_{{x}}): {x} \in W, z_{{x}} \sim Bernoulli(\Phi(\beta S({x})/\sigma)),z_{{x}} = 1\},\nonumber\\
	\tilde{X} &=& \{(\tilde{{x}},z_{\tilde{{x}}}): \tilde{{x}} \in W, z_{\tilde{{x}}} \sim Bernoulli(\Phi(\beta S(\tilde{{x}})/\sigma)),z_{\tilde{{x}}} = 0\}. \nonumber
\end{eqnarray}

The number of points of $W$ is denoted by $k$, of $X$ by $n$ and of $\tilde{X}$ by $k-n$.
$\tilde{X}$ represents the process of the discarded points of $W$ after \textit{thinning}.
The $X$ and $\tilde{X}$ processes are conditionally independent given the
parameters and $\tilde{X}|\tilde{\lambda} \sim PP(\tilde{\lambda})$, with $\tilde{\lambda} =
\lambda^* \ [ 1 - \Phi(\beta S(\tilde{{x}})/\sigma)] = \lambda^* \ \Phi(-\beta S(\tilde{{x}})/\sigma) $.

The likelihood function of the augmented model is given by
\begin{equation}\label{eq:veros}
	l(S,{\theta};{y},{x},w) = \pi({y},{x},{w}|S,{\theta}) = \pi({y}|S,{x},{\theta})\pi({x},{w}|S,{\theta}),
\end{equation}
in which ${w}$ and ${x}$ are the observed locations of $W$ and $X$, respectively, and ${y}$ is the observed values of $Y$.
Note that $W$ is partially observed with $X$ being its observed part and the remainder $\tilde X$
is missing data in the above specification.

The first term of right side of equation (\ref{eq:veros}) is given by
\begin{equation}
	\pi({y}|S,{x},{\theta}) = \left(\dfrac{1}{2\pi\tau^2}\right)^{\frac{n}{2}}  \text{exp}\left\{ -\dfrac{1}{2\tau^2} (y-D\eta - S_n)'(y-D\eta-S_n) \right\}.
\end{equation}
Thus, this term depends on $S = ({S}_n,S_{-n})$ only through ${S}_n$, that represents $S$ at
the $n$ observed locations, and $S_{-n}$ represents the values of process $S$ at all other locations.

The second term of equation (\ref{eq:veros}) is given by $\pi({x},{w}|S,{\theta})$, which is
equivalent to $P({x},{w},k|S,{\theta})$ as follows:
\begin{eqnarray}
	P({x},{w},k|S,{\theta}) &=&  P({x}|{w},S,{\theta})P({w}|k)P(k|\lambda^*), \nonumber\\
	&=&\bigg[ \prod_{i=1}^{n} \dfrac{\lambda(x_i)}{\lambda^*}\bigg] \bigg[
	\prod_{i=n+1}^{k} 1 - \dfrac{\lambda(x_i)}{\lambda^*}\bigg] \bigg[ \prod_{i=1}^{n} \dfrac{1}{|B|}
	\bigg] \text{e}^{-\lambda^*|B|}\dfrac{(\lambda^*|B|)^{k}}{k!}, \nonumber\\
	&=&\Phi_n \left(\dfrac{\beta}{\sigma} S_n;{I}_n\right) \Phi_{k-n}\left(-\dfrac{\beta}{\sigma}
	S_{k-n};{I}_{k-n}\right)\text{e}^{-\lambda^*|B|}\dfrac{(\lambda^*)^k}{k!},  \label{eq:distxw}
\end{eqnarray}
where $\Phi_k(.;{I}_k)$ is the cumulative distribution function of a $k$-dimensional Gaussian
distribution with mean vector ${0}$ and covariance matrix ${I}_k$ and $|B|$ is the volume of region $B$.

Equation (\ref{eq:distxw}) is obtained because $P({x}|{w},S,{\theta})$ is given by the product of the
independent selection (thinning) probabilities $\Phi(\beta S({x})/\sigma)$ for the $X$ locations
whereas the probabilities for the discarded locations of $\tilde X$ are given by the
complementary values $\Phi(-\beta S(\tilde{{x}})/\sigma)$. $P({w}|k)$ is given by a product of the $k$
independent uniform locations of the homogeneous process $W$ over the region of interest $B$
and thus each $w_i$ has density $1/ |B|$. Finally, the distribution of the number $K$ of occurrences
of the homogeneous process $W$ is $Poisson ( \lambda^* |B| )$.

Thus, the likelihood function in (\ref{eq:veros}) depends on $S = ({S}_k,S_{-k})$, only through
${S}_k$. In this way, the integral in (\ref{eq:poisson_process_likel}) is avoided and
the (augmented) likelihood can now be computed and there is
no need for approximations.
Note also that the $\tilde X$ process and its $k-n$ locations are unknown.

The posterior distribution of the unknown components of the model is given by
\begin{eqnarray}
	\pi(S,{\theta}|{y},{x}) &\propto& l({S},{\theta};{y},{x}) \times \pi({S},{\theta}), \nonumber\\
	&\propto& \pi({y}|{S}_n,{x},{\theta})\pi({x},{w},k|{S}_k,{\theta}) \pi(S_{-k}|S_k,\theta) \pi({S}_k|{\theta}) \pi({\theta})\nonumber\\
	&\propto& (\tau^2)^{-n/2} \text{exp}\left\{ -\dfrac{1}{2\tau^2}(y-D\eta-S_n)'(y-D\eta-S_n) \right\} \times \nonumber\\
	&& \Phi_n \left(\dfrac{\beta}{\sigma} S_n;{I}_n\right) \Phi_{k-n}\left(-\dfrac{\beta}{\sigma} S_{k-n};{I}_{k-n}\right)  \text{e}^{-\lambda^*|B|}(\lambda^*)^k \times  \nonumber\\
	&& (\sigma^2)^{-k/2}|{R}|^{-1/2} \text{exp}\left\{ -\dfrac{1}{2\sigma^2} {S}_k'{R}^{-1}{S}_k \right\} \pi(S_{-k}|S_k,\theta) \pi({\theta}), \label{eq:mBA:distpost}
\end{eqnarray}
where $R$ is the correlation matrix for $S_k$ with elements given by the correlation function $\rho$
at each associated pair of locations and $\pi({\theta})$ is the prior distribution of ${\theta}$.

\subsection{Computation}

The posterior distribution of the unknowns from the exact, augmented model is provided (in \ref{eq:mBA:distpost})
and can not be summarized analytically. Posterior summarization may be obtained via MCMC, with
blocks mostly sampled via Gibbs steps. The full conditional distributions for each block can be obtained from (\ref{eq:mBA:distpost}) and are given by

\begin{eqnarray}
	\pi (\lambda^*|.) &\propto& e^{-\lambda^* |B|} (\lambda^*)^k \pi (\lambda^*),  \label{dclambda}\\
	\pi(\tilde{X}|.) &\sim& PP(\tilde{\lambda}), \label{eq:dcxt} \label{eq:ccxtil}\\
	\pi ({S}_k|.) &\propto& \Phi_n \left(\dfrac{\beta}{\sigma} {S}_n;{I}_n\right) \Phi_{k-n}\left(-\dfrac{\beta}{\sigma} {S}_{k-n};{I}_{k-n}\right)\pi({y}|{S}_n,{x},{\theta})\pi({S}_k|{w},{\theta}),\quad  \label{dcsk} \\
	\pi (\eta,\tau^2|.) &\propto& \pi(y|{S}_n,{x},{\theta}) \pi (\eta,\tau^2), \label{dcmutau} \\
	\pi (\sigma^2|.) &\propto& \Phi_n \left(\dfrac{\beta}{\sigma} {S}_n;{I}_n\right) \Phi_{k-n}\left(-\dfrac{\beta}{\sigma} {S}_{k-n};{I}_{k-n}\right) \pi({S}_k|{w},{\theta})\pi(\sigma^2), \label{dcsigma2}\\
	\pi (\phi|.) &\propto&  \pi({S}_k|{w},{\theta}) \pi (\phi),  \label{dcphi}\\
	\pi (\beta|.) &\propto& \Phi_n \left(\dfrac{\beta}{\sigma} {S}_n;{I}_n\right) \Phi_{k-n}\left(-\dfrac{\beta}{\sigma} {S}_{k-n};{I}_{k-n}\right) \pi (\beta), \mbox{ and } \label{dcalphabeta} \\
	\pi (S_{-k}|.) &\propto& \pi(S_{-k}|S_k,\theta).
\end{eqnarray}

The number $k-n$ and the locations of these $k-n$ ocurrences of $\tilde{X}$ are not known and hence become quantities that must
be generated over $B$ in the MCMC scheme,
which means the chain of the discarded points moves through the continuous domain of the region.
The $k$ locations of the augmented process $W$ are obtained by merging the $n$ observed locations of process $X$ with the
simulated locations of $\tilde{X}$. The Gaussian process $S$ may be simulated only on the $k$ points of $W$.
The remaining parameters are sampled after the simulated values of ${x}_k$ and ${S}_k$ are obtained.
Inference for $S_{-k}$ is made through its full conditional posterior based on the values of $S_k$, by retrospective
sampling \citep{papaspiliopoulos2008}. Thus, the following steps shows how to sample from each of the full conditional distributions.

\textbf{Step 1:} $\lambda^*$. If a gamma prior $Gama(a_{\lambda^*},b_{\lambda^*})$ is chosen,
then (\ref{dclambda}) will be a gamma distribution $Gama(a_{\lambda^*}+k,b_{\lambda^*}+|B|)$.

\textbf{Step 2:} $\tilde{X}$. The generation of the locations of the discarded process was made with the following algorithm:\\
\fbox{\parbox{0.95\textwidth}{
		\begin{enumerate}
			\item simulate the number of points $k^*$ from $K^* \sim Pois(\lambda^*|B|)$,
			\item distribute the $k^*$ uniformly over $B$, obtaining the locations $\{\tilde{{x}}_1,...,\tilde{{x}}_{k^*}\}$,
			\item simulate ${S}_{k^*}$ retrospectively from $\pi({S}_{k^*}|{S}_k,{\theta})$,
			\item apply a \textit{thinning} operation with retention probabilities $\Phi(-\beta S(\tilde{{x}}_j)/\sigma)$, $j = 1,...,k^*$,
			\item store the locations retained at step $4$.
		\end{enumerate}
}}\\

Note that the simulation is performed from the conditional distribution of ${S}_{k^*}$
given ${S}_k = \{{S}_n,{S}_{k-n}\}$ (and ${\theta}$), where ${S}_{k-n}$ denotes the Gaussian process
simulated on the $k-n$ points of $\tilde{X}$ in the previous iteration. \\


\textbf{Step 3:} ${S}_k$. The full conditional distribution of ${S}_k$ can be written as the
kernel of a skew normal distribution (SN) (for proof, see the supplementary material).
The form of the SN distribution adopted here is described in the supplementary material. In this way, equation
(\ref{dcsk}) can be written as
\begin{eqnarray}
	\pi ({S}_k|.) &\propto& \phi_k({S}_k;{\mu}^*,{\Sigma^*})\Phi_k ({G} {S}_k;{I}_K), \label{eq:skskewn}
\end{eqnarray}
where
\begin{eqnarray}
	{\Sigma^*} &=& ({C}'{{\Sigma}_y^{-1}}{C} + (\sigma^2{R})^{-1})^{-1}, \nonumber\\
	{\mu}^* &=& {\Sigma^*}C'{{\Sigma}_y^{-1}}({y}-D\eta), \nonumber\\
	{G} &=& (\beta/\sigma ) * \text{diag}({I}_n,-{I}_{k-n}),\nonumber
\end{eqnarray}
${\Sigma}_y = \tau^2{I}_N$ and ${C}$ is the $n\times k$ matrix whose $i$th row consists of
$n-1$ zeros and a single 1 identifying the $i$-th position in the ${S}_k$ vector,  $i = 1,...,n$.
This is the kernel of a $SN({\mu}^*,{\Sigma^*},{G})$ distribution that can be sampled by
Algorithm 4 in \cite{geg2018}.\\

\textbf{Step 4:} ${\eta}$ and $\tau^2$. It was assumed that, a priori, ${\eta}$ is independent
of $\tau^2$ and $\pi({\eta},\tau^2) = \pi({\eta})\pi(\tau^2)$, where $\eta\sim N_{p+1}(\eta',\sigma^2_\eta I_{p+1})$,
$\eta' = (\eta'_0,\eta'_1,...,\eta'_p)$, and $\tau^2 \sim IG(a_\tau,b_\tau)$, where $IG$ is the inverse Gamma distribution. In this way, the full conditional distribution of $\eta$ is given by $(\eta|.) \sim N_{p+1}(\mu^*_\eta, \Sigma^*_\eta)$, where
\begin{eqnarray}
	\mu^*_\eta &=& \Sigma^*_\eta \left( \dfrac{1}{\tau^2} D'(y-S_n) + \dfrac{1}{\sigma^2_\eta}\eta' \right), \nonumber\\
	\Sigma^*_{\eta} &=& \left( \dfrac{1}{\tau^2} D'D + \dfrac{1}{\sigma^2_\eta}I_{p+1} \right). \nonumber
\end{eqnarray}

The full conditional distribution of $\tau^2$ is
\begin{eqnarray}
	(\tau^2|.) &\sim& GI \bigg( \dfrac{n}{2} + a_\tau, \dfrac{\sum_{i=1}^{n}(y_i - S({x}_i) - D\eta)^2 }{2} + b_\tau \bigg).\nonumber
\end{eqnarray}

\textbf{Step 5:} $\sigma^2$. A MH step was implemented for sampling this parameter. If a prior distribution $IG(a_{\sigma},b_{\sigma})$ is assumed, the full conditional distribution is given by
\begin{eqnarray}
	\pi (\sigma^2|.) &\propto& \Phi_n \left(\dfrac{\beta}{\sigma} {S}_n;{I}_n\right) \Phi_{k-n}\left(-\dfrac{\beta}{\sigma} {S}_{k-n};{I}_{k-n}\right) \times \nonumber\\
	&& \left( \dfrac{1}{\sigma^2}\right)^{k/2+a_{\sigma} +1}  \text{exp}\left\{- \left(  \dfrac{{S}_k'{R}^{-1}{S}_k}{2}  + b_{\sigma}  \right) \dfrac{1}{\sigma^2}  \right\}. \nonumber
\end{eqnarray}

Considering the proposal distribution $ q(\sigma^2_p|\sigma^2_c)$ as $\text{Lognormal}(\text{log}(\sigma^2_c); \delta_{\sigma^2})$, where $\sigma^2_p$ is the proposed value for $\sigma^2$ and $\sigma^2_c$ is the current value of the chain, the acceptance probability is $A_{\sigma} = \text{min}\{1,p_{\sigma}\}$, where
\begin{eqnarray}
	p_{\sigma} = \dfrac{\Phi_k \big((\beta/\sigma_p) {S}_k{I}_k^*;{I}_k\big)}{\Phi_k \big((\beta/\sigma_c) {S}_k{I}_k^*;{I}_k\big)} \left(\dfrac{\sigma^2_p}{\sigma^2_c}\right)^{-\frac{k}{2}-a_{\sigma}} \text{exp}\left\{- \left(  \dfrac{{S}_k'{R}^{-1}{S}_k}{2}  + b_{\sigma}  \right) \left( \dfrac{1}{\sigma^2_p} - \dfrac{1}{\sigma^2_c} \right)   \right\}.\nonumber
\end{eqnarray}

\textbf{Step 6:} $\phi$.  A MH step was implemented for sampling this parameter. In the case is a univariate component, say the range,
adopting a prior distribution $Gamma(a_\phi,b_\phi)$, the full conditional distribution is written as
\begin{equation}\label{eq:dccphi}
	\pi(\phi|.) \propto |{R}|^{-1/2} \phi^{a_\phi -1} \text{exp}\bigg\{ -\dfrac{1}{2\sigma^2}{S}_k'{R}^{-1}{S}_k - b_\phi \phi \bigg\}.\nonumber
\end{equation}

We considered a proposal distribution $q(\phi_p|\phi_c)$ as $\text{Lognormal}(\text{log}(\phi_c); \delta_\phi)$, where $\phi_p$ is the proposed value for $\phi$ and $\phi_c$ is the current value. The acceptance probability is $A_\phi = \text{min}\{1,p_\phi\}$, where
\begin{eqnarray}
	p_\phi = \left( \dfrac{|{R}(\phi_p)|}{|{R}(\phi_c)|} \right) ^{-1/2} \left(\dfrac{\phi_p}{\phi_c}\right) ^{a_\phi}
	\text{exp} \bigg\{ -\dfrac{1}{2\sigma^2}\Big({S}_k'({R}(\phi_p)^{-1} - {R}(\phi_c)^{-1}){S}_k \Big) - b_\phi(\phi_p- \phi_c)\bigg\}.\nonumber
\end{eqnarray}
Straightforward generalization is performed for muitivarite $\phi$ along the same lines as those detailed above.

\textbf{Step 7:} $\beta$. A MH step was implemented for sampling this parameter. Considering a prior distribution $N(\mu_\beta,\sigma^2_\beta)$, the full conditional distribution is given by
\begin{eqnarray}
	\pi (\beta|.) &\propto& \Phi_n \left(\dfrac{\beta}{\sigma} {S}_n;{I}_n\right) \Phi_{k-n}\left(-\dfrac{\beta}{\sigma} {S}_{k-n};{I}_{k-n}\right) \phi (\beta;\mu_\beta,\sigma^2_\beta). \nonumber
\end{eqnarray}

Assuming the proposal distribution of $\beta_p|\beta_c$ as $N(\beta_c,\delta_\beta)$, where $\beta_p$ is the proposed value for $\beta$ and $\beta_c$ is the current value, the acceptance probability is $A_\beta = \text{min}\{1,p_\beta\}$, where
\begin{eqnarray}
	p_\beta &=& \dfrac{\Phi_n \left(\dfrac{\beta_p}{\sigma} {S}_n;{I}_n\right) \Phi_{k-n}\left(-\dfrac{\beta_p}{\sigma} {S}_{k-n};{I}_{k-n}\right) \phi (\beta_p;\mu_\beta,\sigma^2_\beta)}{\Phi_n \left(\dfrac{\beta_c}{\sigma} {S}_n;{I}_n\right) \Phi_{k-n}\left(-\dfrac{\beta_c}{\sigma} S_{k-n};{I}_{k-n}\right) \phi (\beta_c;\mu_\beta,\sigma^2_\beta)}. \nonumber
\end{eqnarray}

\textbf{Step 8:} $S_{-k}$. Any finite subset of $S^*_{-k}$ at the $x^*$ locations is easily
sampled by kriging. Note that since $S$ is a Gaussian process, the term $({S}^*_{-k},{S}_k|{\theta})$
is normally distributed with
\begin{equation*}\label{eq:s0sk}
	({S}^*_{-k},{S}_{k}|{\theta}) \sim N_{(n^* + k)} \left( \left(
	\renewcommand{\arraystretch}{1}
	\begin{array}{c}
		{0}\\
		{0}
	\end{array} \right) , \sigma^2 \left(
	\renewcommand{\arraystretch}{1}
	\begin{array}{cc}
		{R}_{11} & {R}_{12}\\
		{R}_{21} & {R}_{22}\\
	\end{array} \right) \right),
\end{equation*}
in which $n^*$ is the number of elements of $S^*_{-k}$, ${R}_{11}$ is the covariance matrix
of ${S}^*_{-k}$, ${R}_{22}$ is the covariance matrix of ${S}_k$, ${R}_{12}$ is the covariance
matrix of ${S}^*_{-k}$ and ${S}_k$ and ${R}_{21} = {R}'_{12}$. In this way,
$({S}^*_{-k}|{S}_k,{\theta}) \sim  N_{n^*}({\mu}_{{S}^*_{-k}|{S}_k},
$\quad$\sigma^2{R}_{{S}^*_{-k}|{S}_k})$, where ${\mu}_{{S}^*_{-k}|{S}_k} = {R}_{12}{R}_{22}^{-1}{S}_k$
and ${R}_{{S}^*_{-k}|{S}_k} = {R}_{11} - {R}_{12}{R}_{22}^{-1}{R}_{21}$.

The steps 1-7 will provide, after empirical convergence of the Markov chains, approximate
samples of the posterior distribution of the unknown quantities of the model. From these samples,
it is possible to make inference for the parameters of the model.

Prediction of the response $Y$ at unobserved locations is the primary goal of many geostatistical analyses.
The next section will provide the algorithm for drawing samples of the response variable $Y$ at unobserved locations.
Step 8 above is crucial for making prediction of the response at these locations and is used to obtain predictive samples.

\subsection{Prediction}
One of the main objectives of geostatistics is to make prediction for the variable of interest
$Y$ at unobserved locations. In some geostatistical analysis, a regular grid is defined
over the region $B$ for construction of maps.
Let $x_u = (x_{u_1},x_{u_2},...,x_{u_{n_u}})$ be the vector of unobserved locations of interest,
defined irrespective of whether the chosen grid is regular or irregular. Also, define
$S_u$ and $Y_u$ as the values of $S$ and $Y$ at $x_u$.

The posterior predictive distribution of $Y_u$ is given by
\begin{eqnarray}
	\pi({Y}_u|{x},{y}) &=& \int \pi({Y}_u,{S}_u,{S}_k,{\theta}|{x},{y}) d{S}_u d{S}_k d{\theta} \nonumber\\
	&=& \int \pi({Y}_u|{S}_u,{\theta}) \pi({S}_u|{S}_k,{\theta}) \pi({S}_k,{\theta}|{x},{y}) d{S}_u d{S}_k d{\theta},\label{eq:mBA:pred:Yp}
\end{eqnarray}
where $\pi({S}_k,{\theta}|{x},{y})$ is the posterior distribution of $({S}_k,{\theta})$.
Note that the posterior distribution of $S_u$ given $S_k$ and $\theta$ is obtained by kriging,
as described in the step 8 of the MCMC algorithm above and a sample of $S_u$ is easily drawn. Then,
a sample of $Y_u$ is obtained by  equation (\ref{eq:mBA:Y}) of the model. Thus, any estimate of
$Y_u$, say $\hat{Y}_u$, such as the mean or the median of the distribution can be calculated.
It can also be approximated by Monte Carlo sampling.

Another interesting quantity to predict is the intensity function at the unobserved locations.
Let $\lambda(x_u) = (\lambda(x_{u_1}),\lambda(x_{u_2}),...,\lambda(x_{u_{n_u}}))'$ be the vector
of the values of the intensity function at $x_u$. A posterior sample of $\lambda(x_u)$ is easly obtained
by replacing samples of $S_u$, $\lambda^*$ and $\theta$ in (\ref{eq:mBA:lambda}).
This task requires posterior samples of $S$ at $x_u$ and $\theta$. Thus, for each posterior sample of
$(S_k,\lambda^*,\theta)$, $S_u$ is simulated via retrospective sampling and $\lambda(x_u)$ can then be calculated.

\subsection{Further information about the model}

It is well known that the correlation parameters $\phi$ are difficult to estimate \citep{diggle2007},
even in the simple case of the exponential correlation function $\rho(h) = \exp\{-h/\phi\}$.
In this way, an informative prior or even a fixed value are sometimes assumed. We rely on the study
of \cite{paez2005} where different priors for the correlation parameter were considered and compared.
The parameter was mostly estimated with large uncertainty, but the best results were
obtained with vaguely informative priors, with informed means (merely in terms of the order of magnitude) and
large variances. This type of prior is adopted here.

An extension of the proposed model is the inclusion of an intercept in the link function of
equation (\ref{eq:mBA:lambda}). The intensity function thus becomes
\begin{eqnarray}
	\lambda(x) &=&  \lambda^* \Phi(\alpha + \beta S(x)/\sigma). \label{eq:mBA:lambda2}
\end{eqnarray}
This specification generates an identifiability problem between $\lambda^*$ and $\alpha$ even tough
it could provide more flexibility to the model. Both parameters are related to the number of points of the
Poisson process and are associated with the overall magnitude of the intensity of the process. As a result,
$\alpha$ is poorly estimated unless a highly informative prior is assumed.
This issue is well discussed in \cite{geg2018}, and the authors propose possible alternatives to
bring in identifiability. In the context of this paper, the model as initially proposed without the intercept $\alpha$ seems to
provide model identification and does not restrict model capabilities.
Possible lack of fit from removal of $\alpha$ is compensated by the remaining parameters.
This strategy is retained here.

\section{Simulation study}\label{sec:simul}
In this section, the results of a simulation study are presented in order to evaluate the
performance of the proposed methodology. The main findings are shown below and the remainder is presented in the supplementary material. All computations
were performed in the \texttt{R} software \citep{r2020} integrated with the \texttt{C++} language
via \texttt{Rcpp} \citep{rcpp}.

\subsection{Data with preferential sampling}\label{sec:simul:ps}
Simulations were performed over the unit square $B= [0,1]^2 \subset R^2$ as the region of interest
and the true values of the parameters were assumed to be $(\lambda^*,\mu,\tau^2,\sigma^2,\phi,\beta)
= (150,4,0.10,3,0.15,2)$ and the exponential correlation function was assumed.

The following prior distributions were adopted:
$\lambda^* \sim Gamma(0.001, 0.001)$, $\mu \sim N(0,10^6)$, $\tau^2 \sim IG(0.001,0.001)$,
$\sigma^2 \sim IG(0.001,0.001)$, $\phi \sim Gamma(2,4)$, and $\beta \sim N(0,1)$.
They are mostly providing little information about the parameters, as a representation
of vague prior information.

\begin{figure}[b!]
	\centering
	\subfigure[]{
		\includegraphics[scale=0.31]{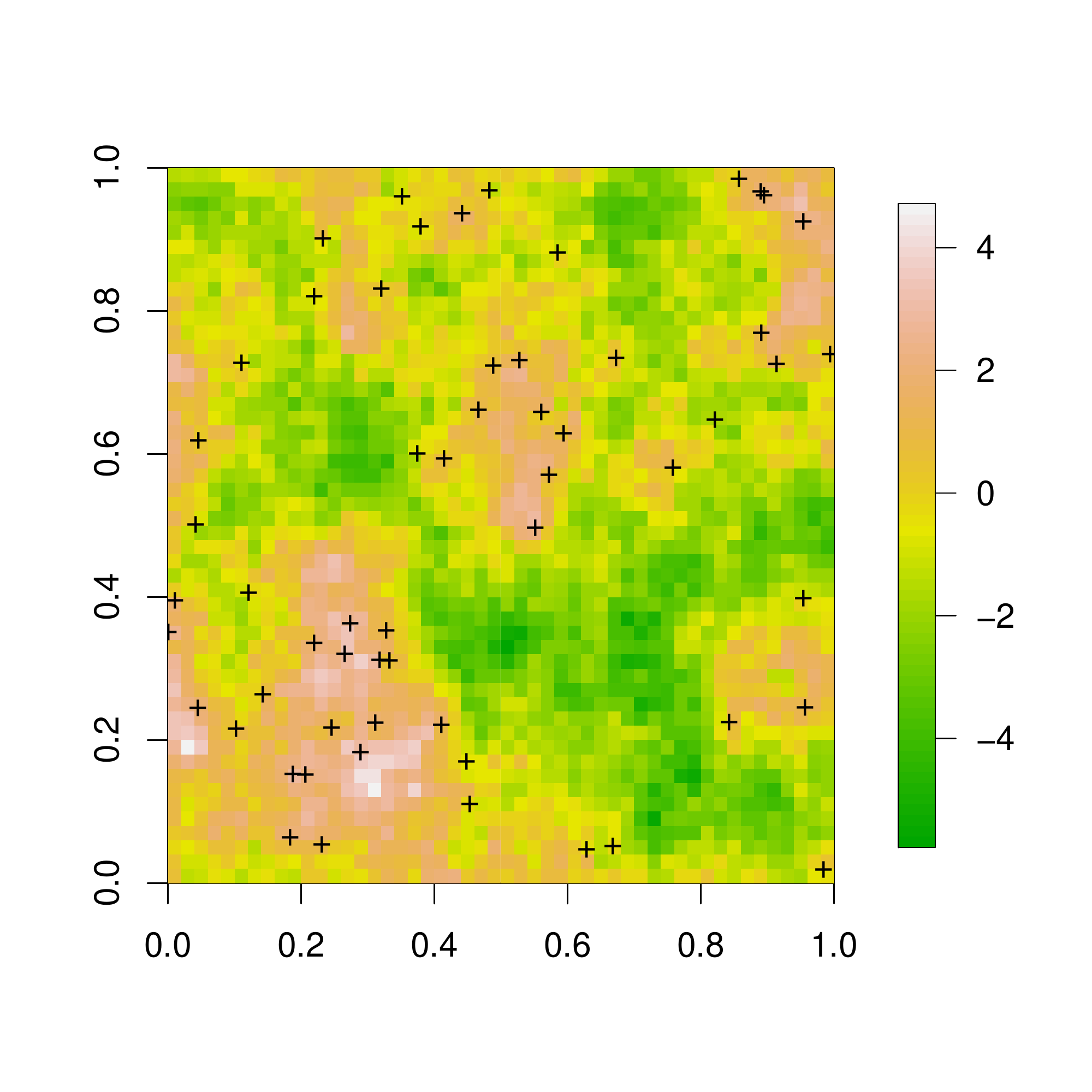}
	}
	\subfigure[]{
		\includegraphics[scale=0.31]{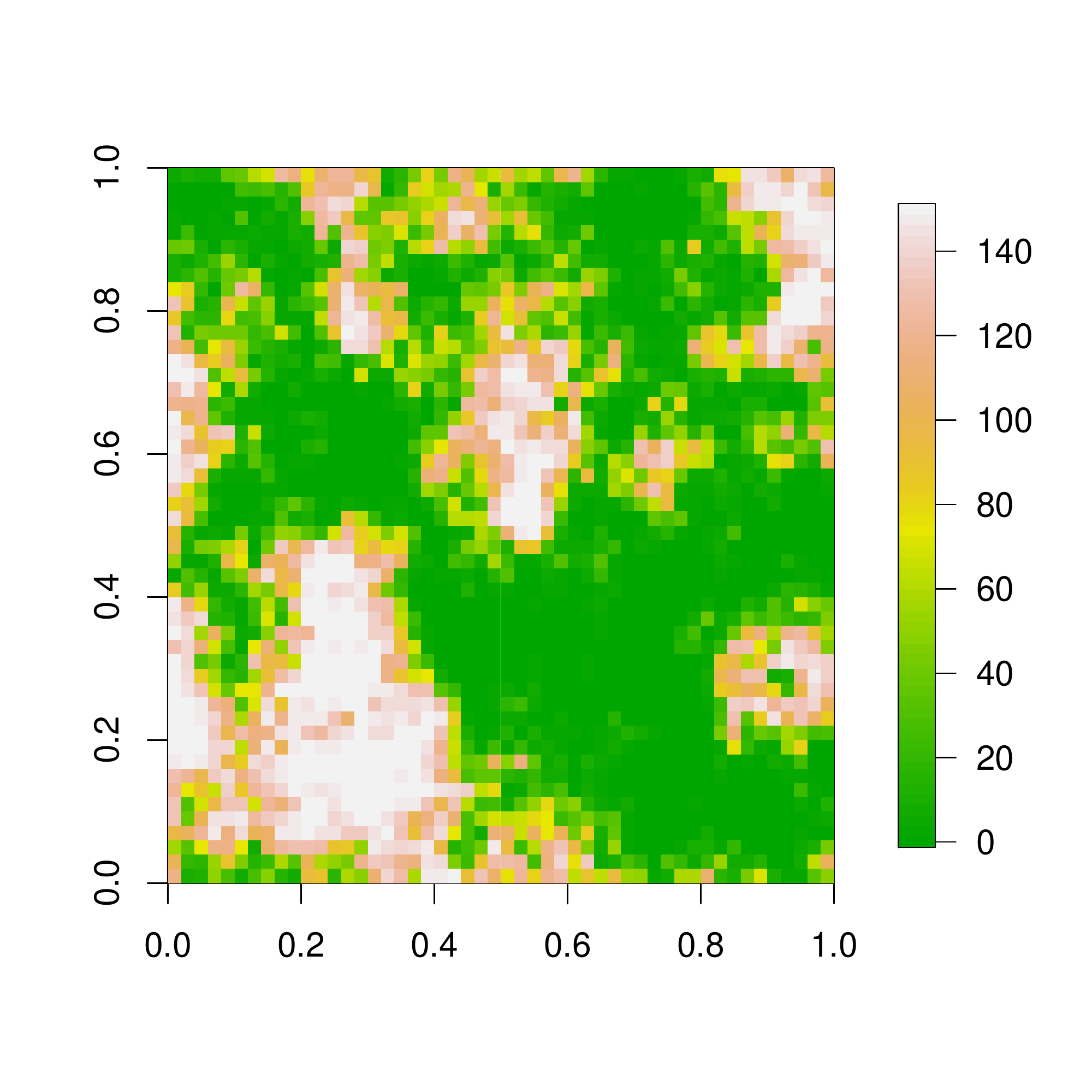}
	}
	\caption{Maps of (a) the Gaussian process and (b) the intensity function of simulated
		dataset 1 considering a preferential sampling generating scheme. \label{fig:simul:1:mapa:sn_gauss}}
\end{figure}

Ten datasets were simulated and the parameters were estimated for each one of them. The sizes of the
samples are showed in Table \ref{tab:simul:ps}. As an example, Figure \ref{fig:simul:1:mapa:sn_gauss}
presents the maps of the Gaussian process and the intensity function over $B$ for dataset 1.

\begin{table}[b!]
	\centering \renewcommand\arraystretch{1.2}
	\begin{tabular}{lcccccccccc}
		\hline
		Dataset&1&2&3&4&5&6&7&8&9&10\\ \hline
		n&58&63&89&72&85&80&75&108&55&87\\
		\hline
	\end{tabular}
	\caption{Size of the simulated datasets with preferential sampling.}\label{tab:simul:ps}
\end{table}

The MCMC algorithm was run for 200,000 iterations with a burn-in of size 20,000. The posterior
sample was constructed with a lag of 60, resulting in a 6,000 sample size. The convergence of the
Markov chains was verified graphically through the posterior parameters trace plots (the plots are
shown in the supplementary material). The exact model with preferential sampling will be denote by
EPS, the discrete model with preferential sampling by DPS and the non preferential model by NPS.

\begin{figure}[hbt!]
	\centering
	\includegraphics[scale=0.32]{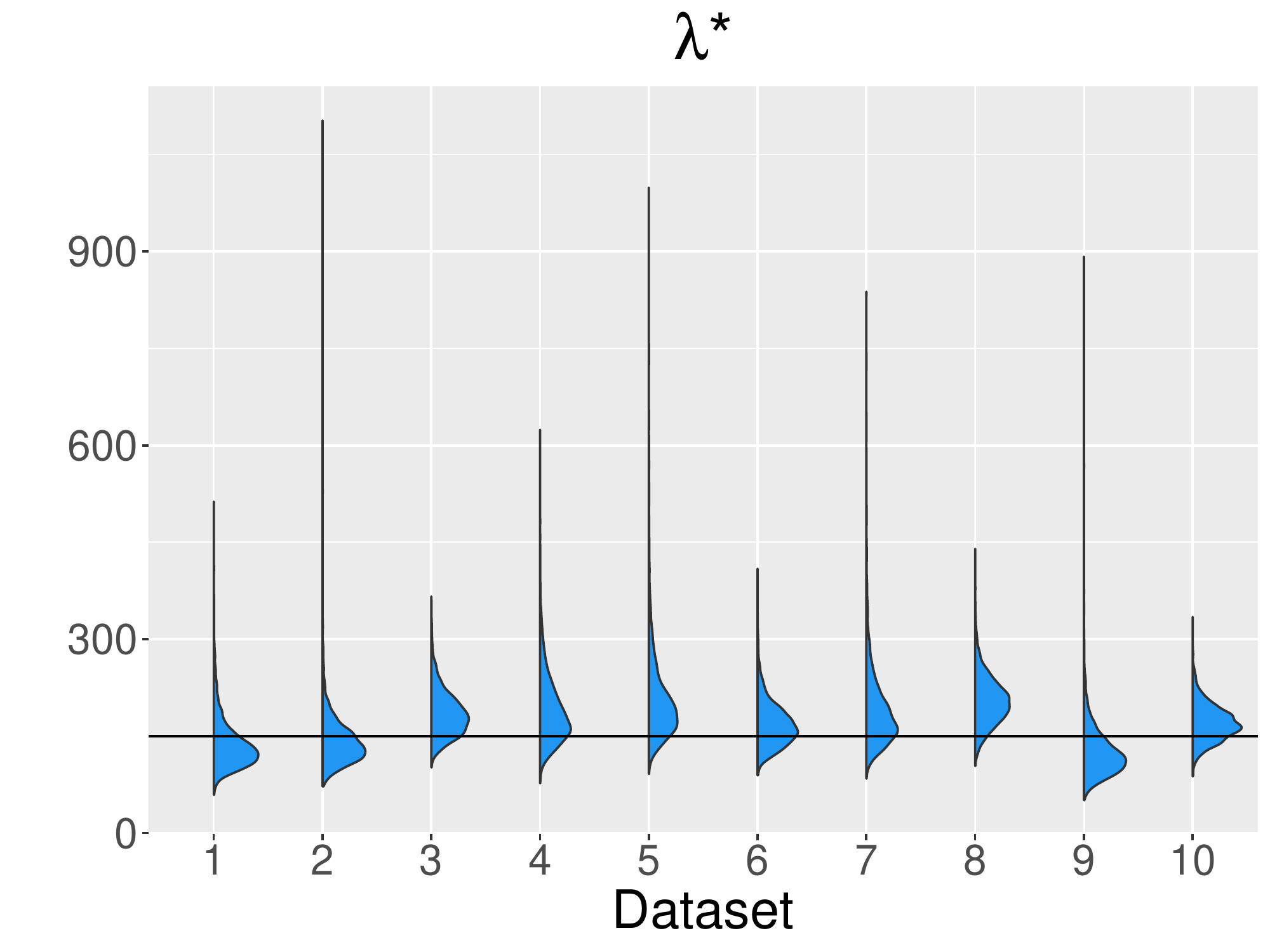}
	\includegraphics[scale=0.32]{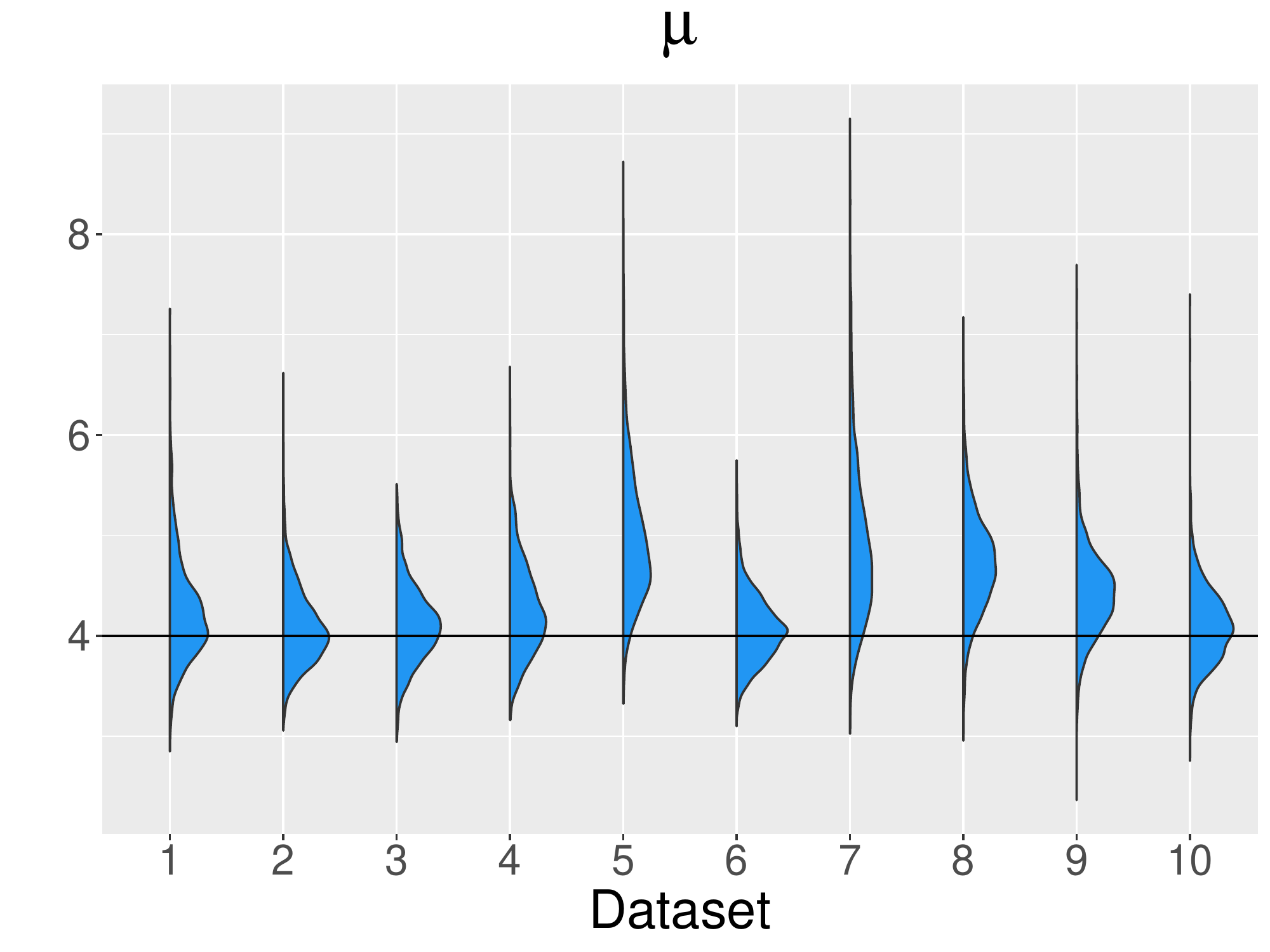}\\
	\includegraphics[scale=0.32]{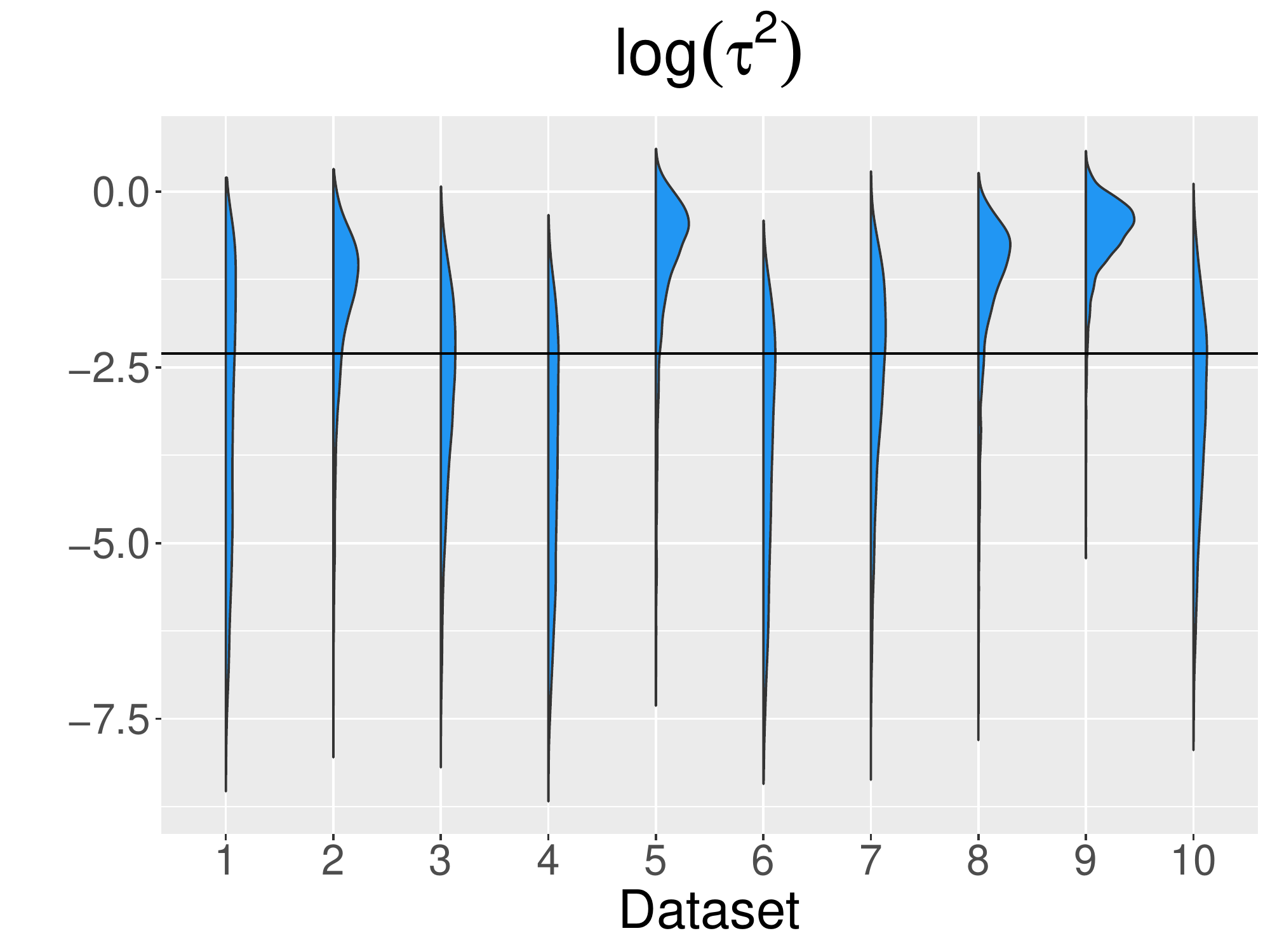}
	\includegraphics[scale=0.32]{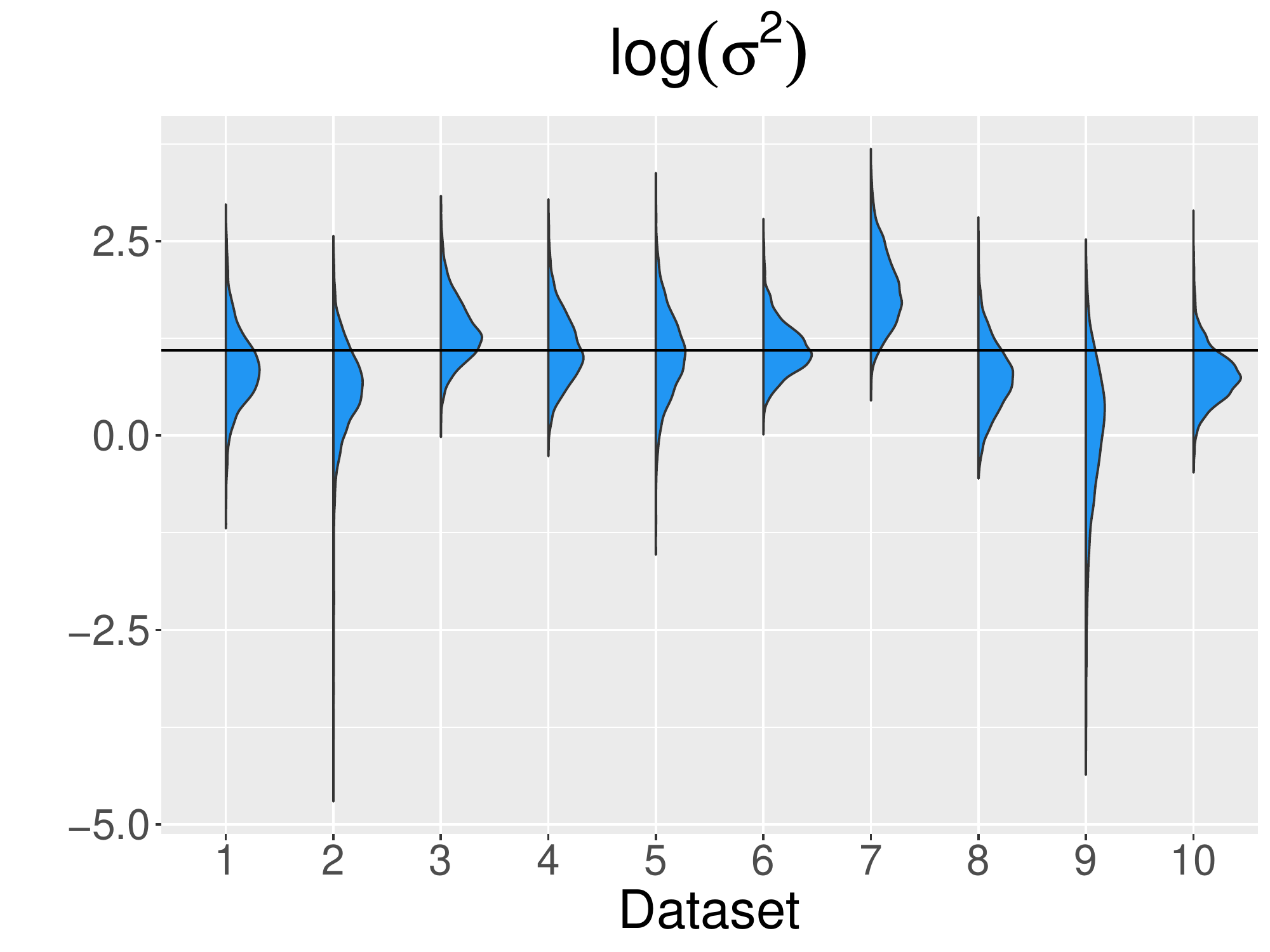}\\
	\includegraphics[scale=0.32]{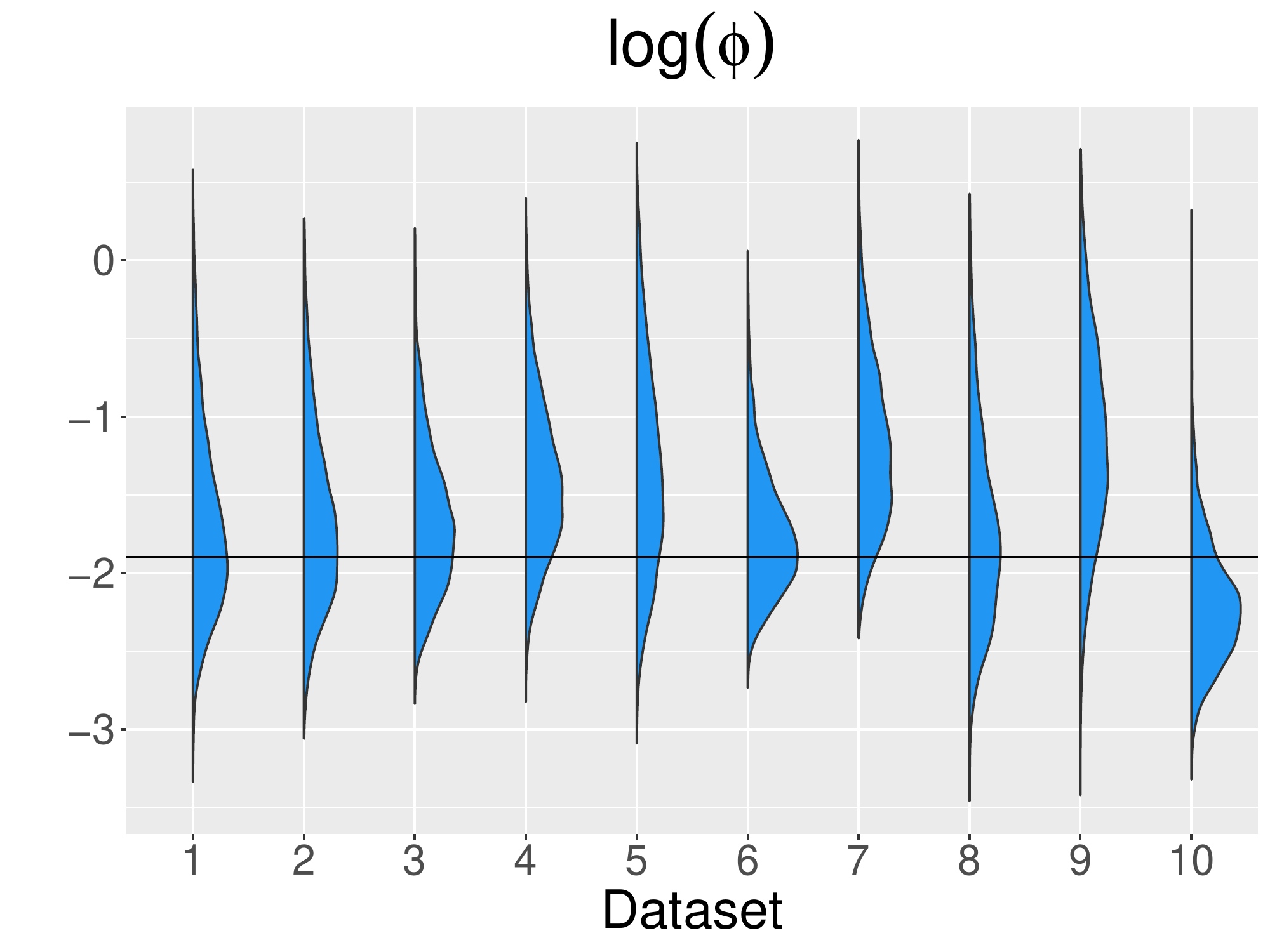}
	\includegraphics[scale=0.32]{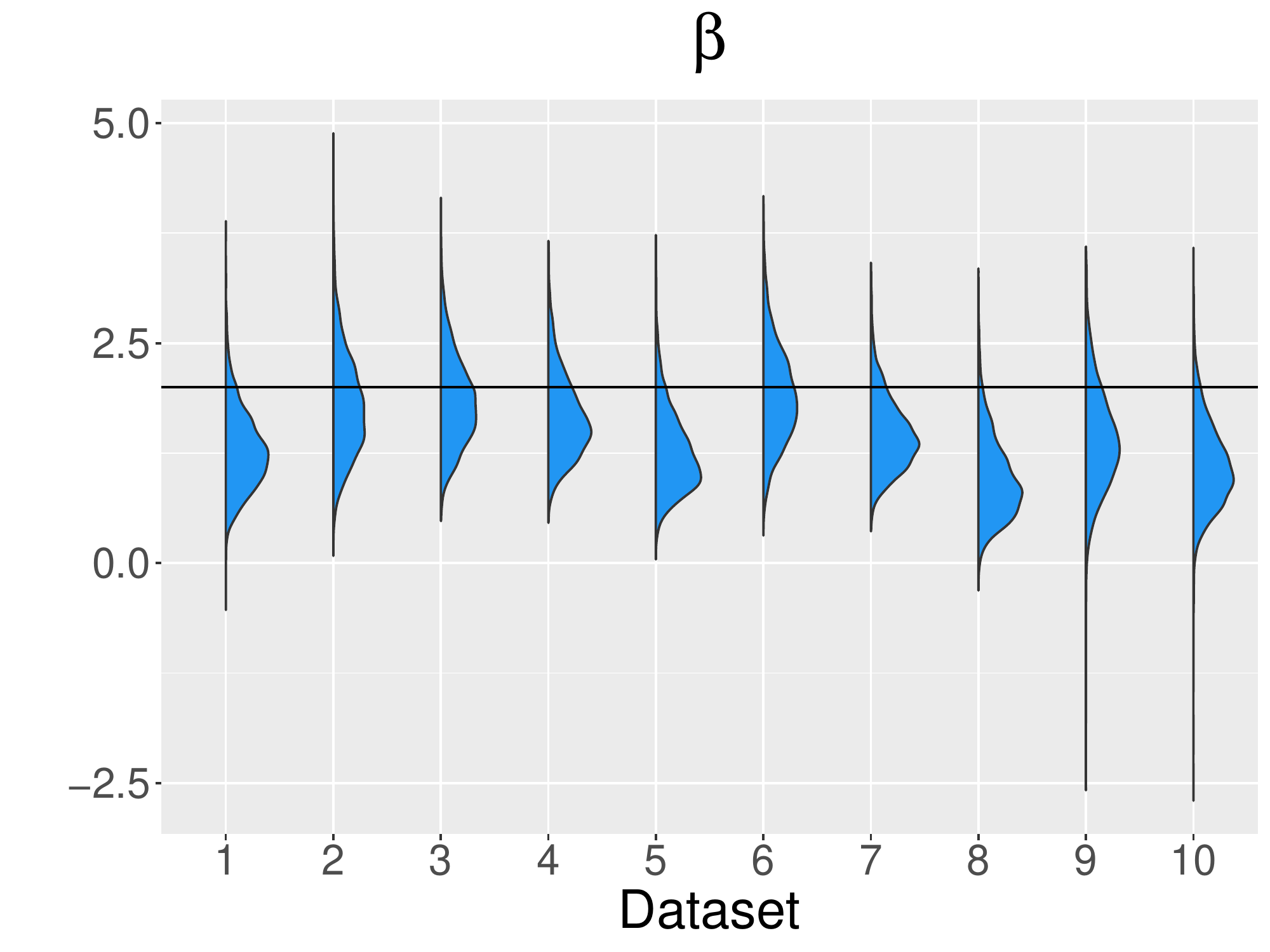}
	\caption{Posterior densities of the parameters of the EPS model for data simulated with preferential sampling. The horizontal line represents the true value of the parameter. \label{fig:simul:bayes:sec1:hist:10:beta2}}
\end{figure}

Figure  \ref{fig:simul:bayes:sec1:hist:10:beta2} shows the posterior densities of the parameters.
Good parameter estimation is observed for most of the datasets, for all parameters, despite their modest
sample sizes. The $\beta$
parameter was recovered well in half of the datasets and was underestimated in the rest, but
showing a very substantial probability mass at positive values for all datasets. This means that
the model was able to capture the sampling preferentiality present in the synthetic datasets.

\begin{figure}[hbt!]
	\centering
	\includegraphics[scale=0.32]{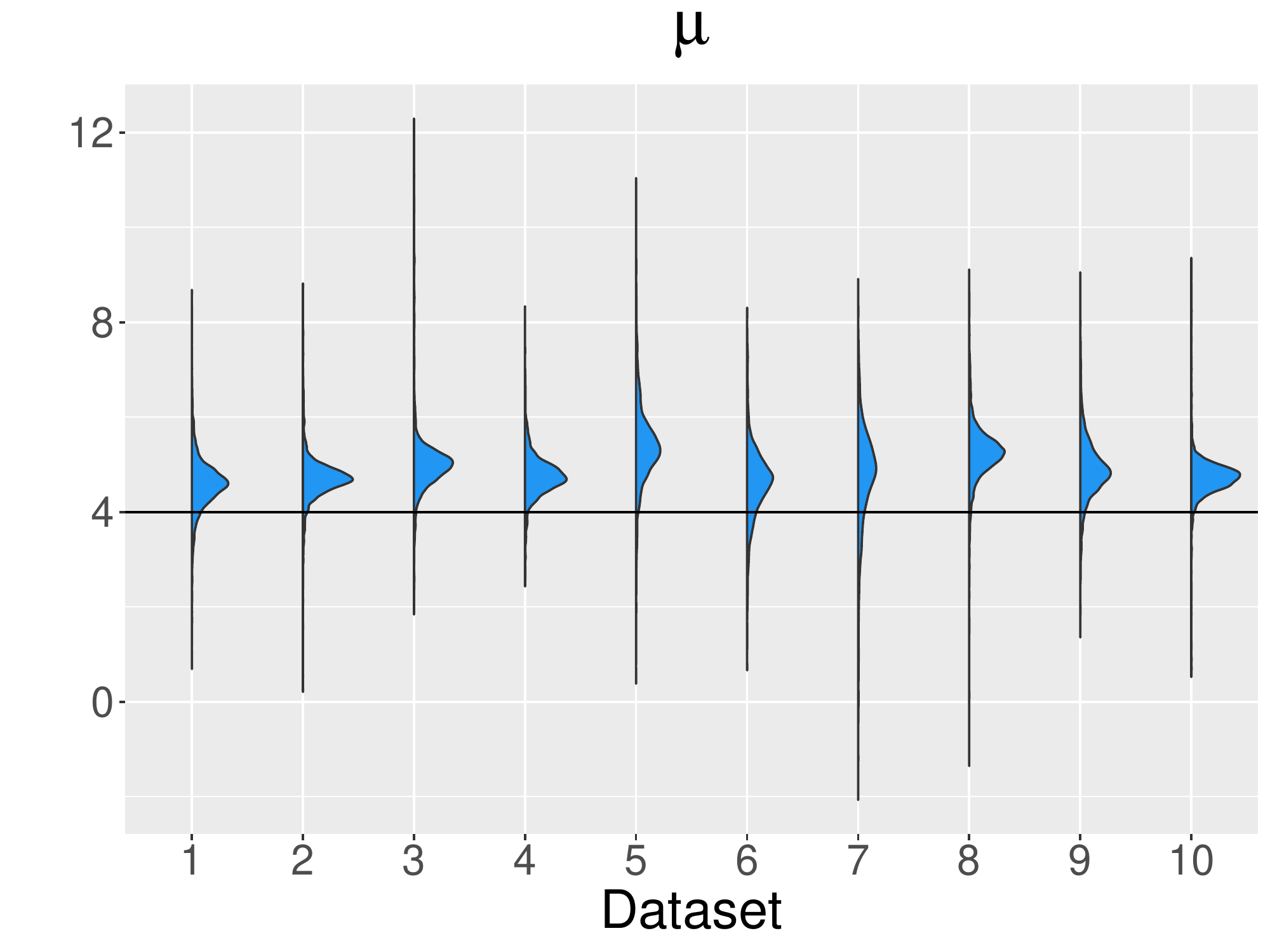}
	\includegraphics[scale=0.32]{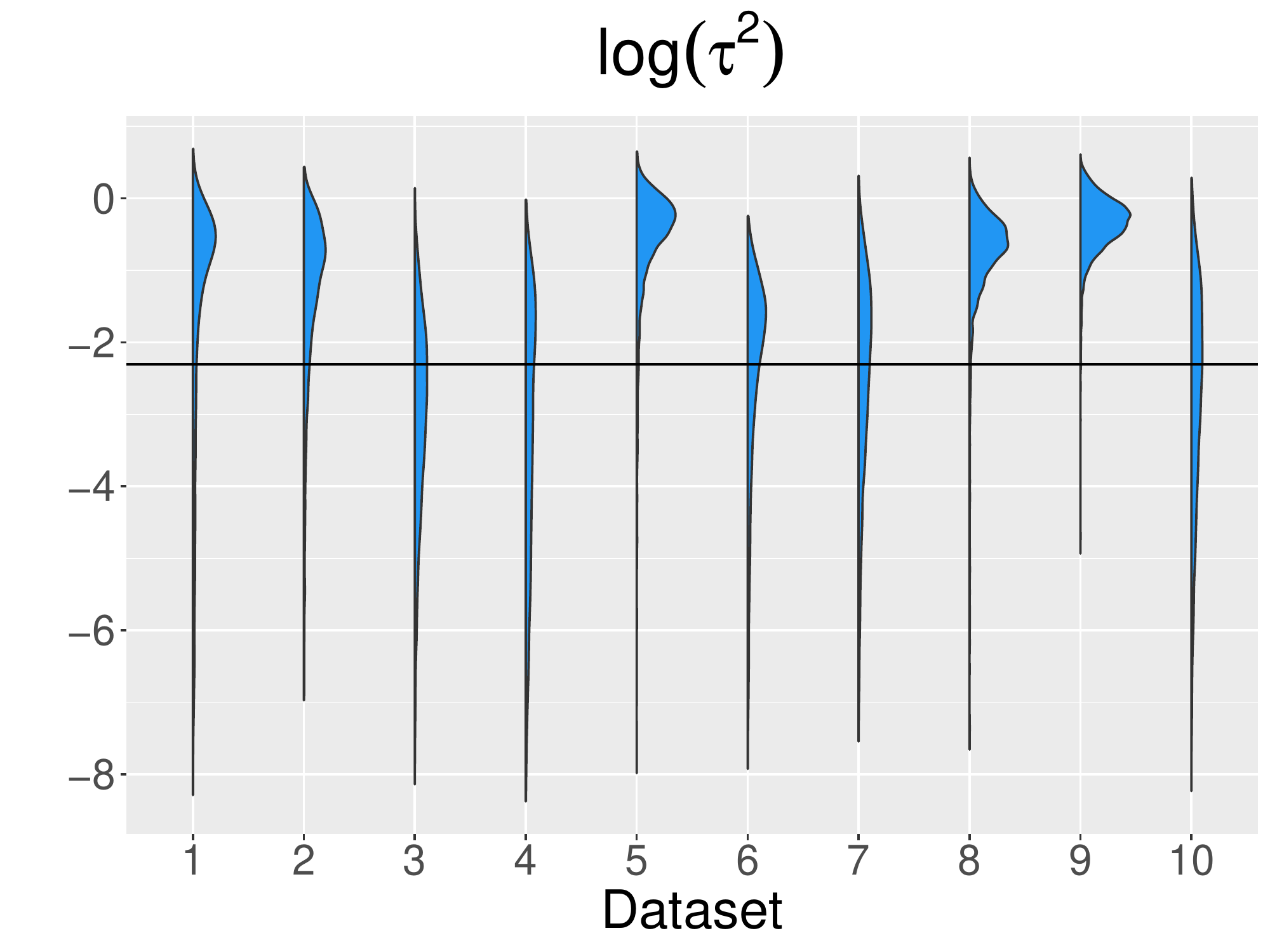}\\
	\includegraphics[scale=0.32]{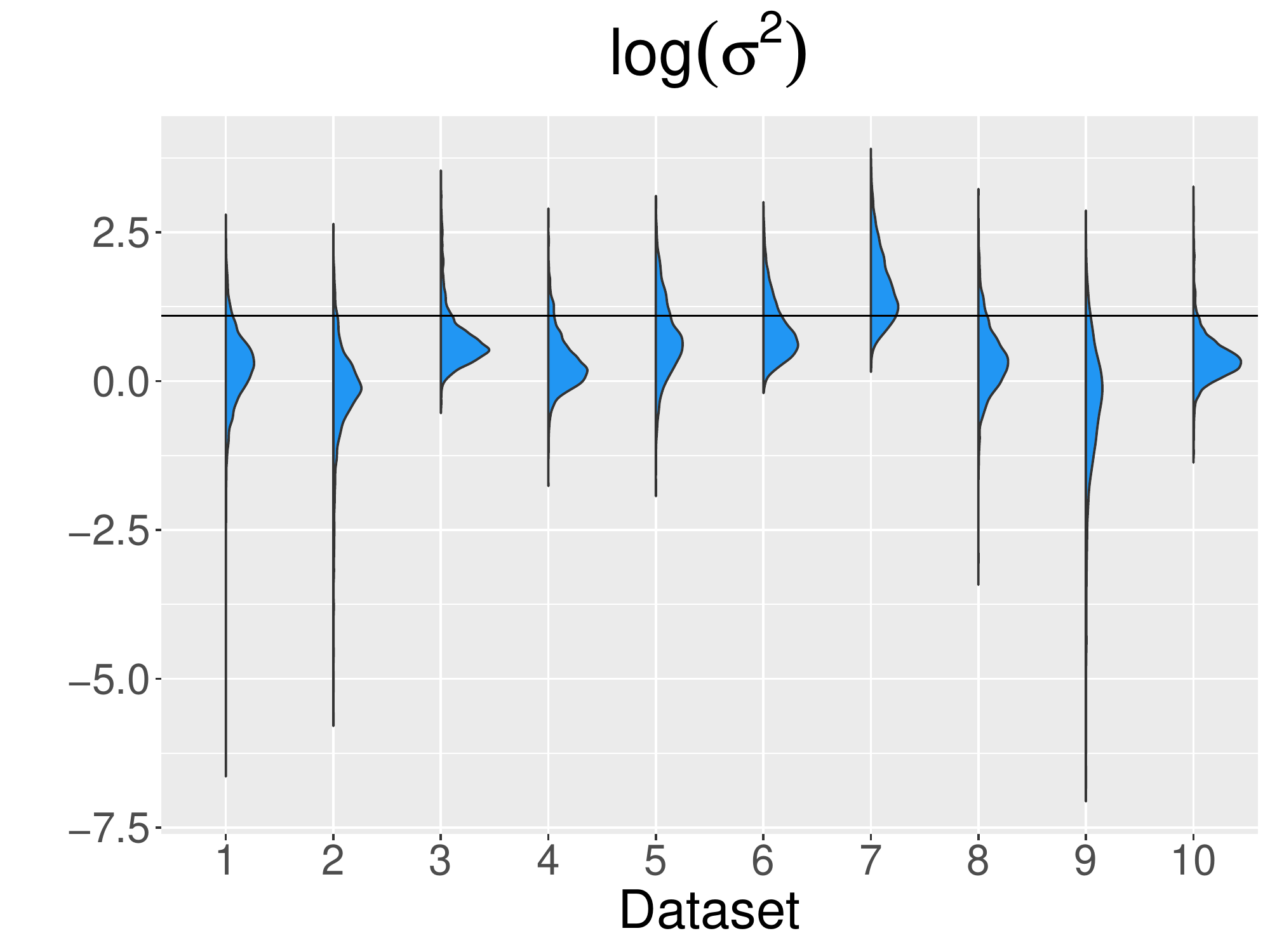}
	\includegraphics[scale=0.32]{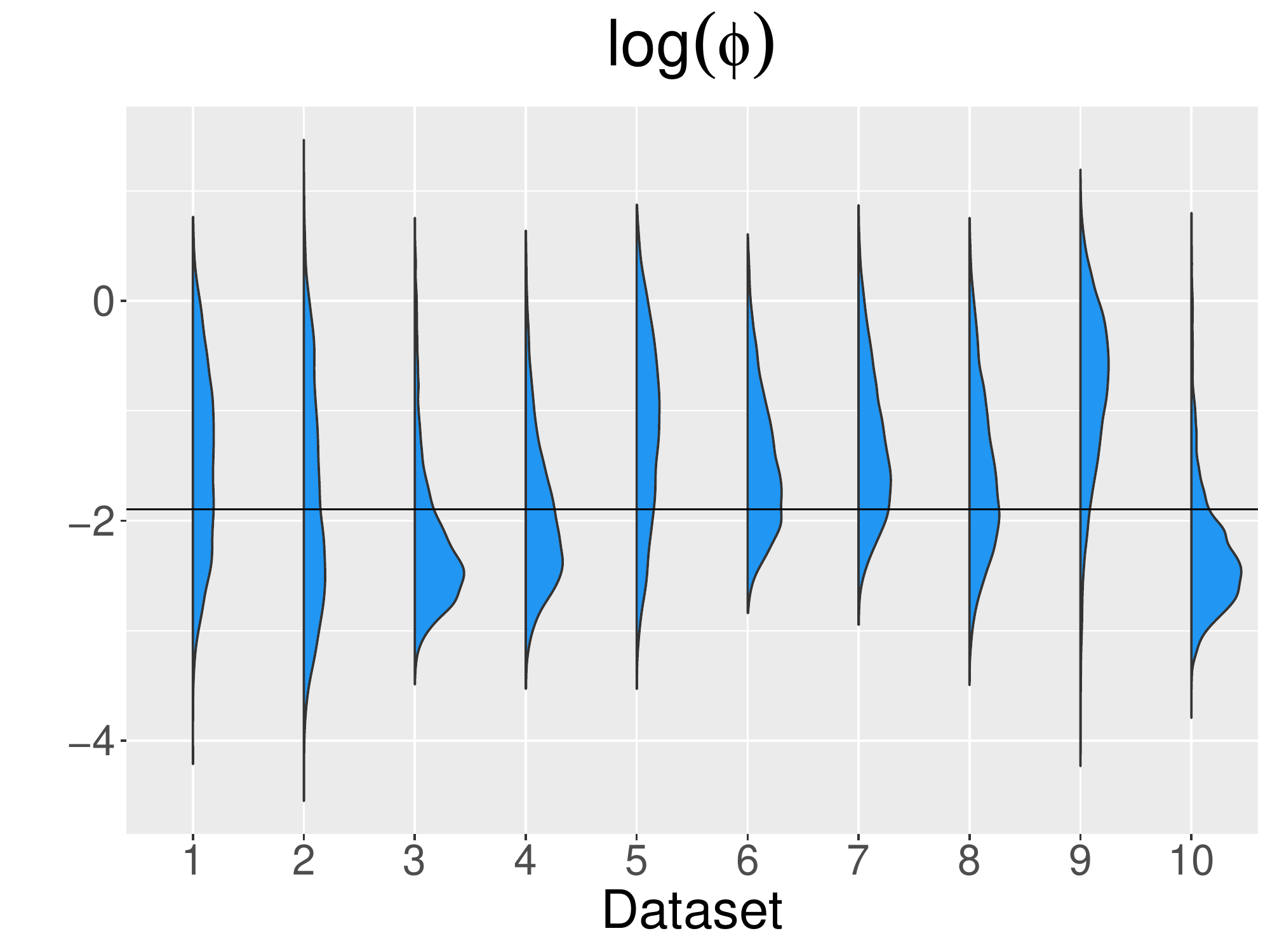}
	\caption{Posterior densities of the parameters of the NPS model for data simulated with preferential sampling. The horizontal line represents the true value of the parameter. \label{fig:simul:bayes:sec1:hist:10:sbeta2}}
\end{figure}

The parameters were also estimated using the non preferential model and results are showed in
Figure \ref{fig:simul:bayes:sec1:hist:10:sbeta2}. The mean parameter is overestimated in all
datasets, as expected, to compensate for the lack of preferentiality in the model.

Two measures were considered to assess the predictive performance of the models:
the mean absolute prediction error (MAPE) and the coverage ratio of credibility intervals (CRCI),
given by
\begin{eqnarray}
	MAPE  &=&  \frac{1}{n_p}\sum_{i=1}^{n_p}|\hat{Y}(x_i) - y(x_i)|,\nonumber\\
	CRCI_\alpha
	&=&  \frac{1}{n_p}\sum_{i=1}^{n_p}\mathbbm{1}\{y(x_i)\in CI(Y(x_i),\alpha )\},\nonumber
\end{eqnarray}
where $CI( Z , \alpha )$ stands for a $( 1 - \alpha )$ predictive credibility
interval for $Z$, $\hat{Y}(x_i)$ and $y(x_i)$ are the respective
mean of the predictive distribution and observed value of $Y$ at the unobserved location $x_i$ and $n_p$ is the size of the vector of unobserved locations for prediction, for
$i = 1,...,n_p$. Better models should present CRCI levels closer to the nominal level $1 - \alpha$.

\begin{figure}[hbt!]
	\centering
	\includegraphics[scale=0.25]{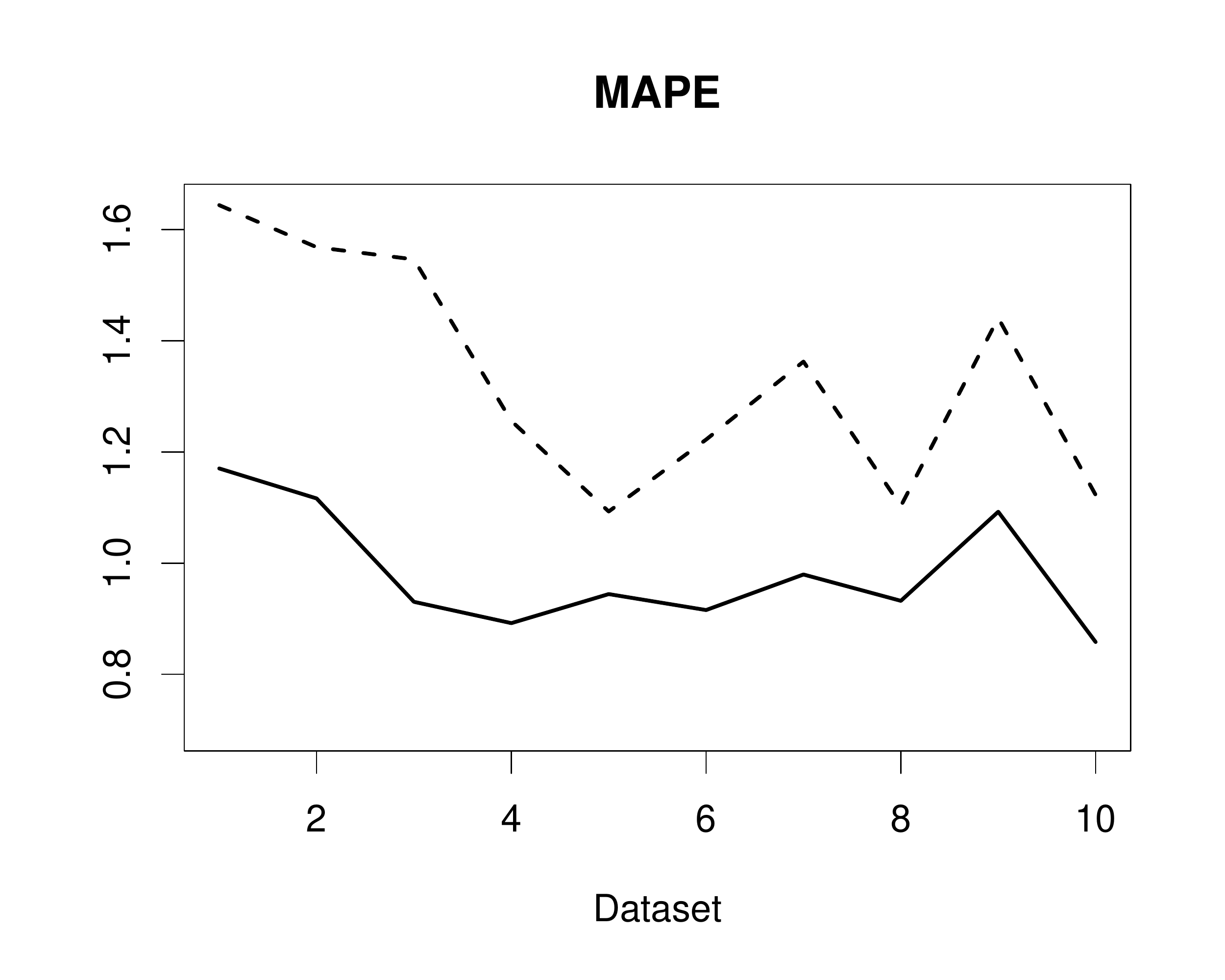}
	\includegraphics[scale=0.25]{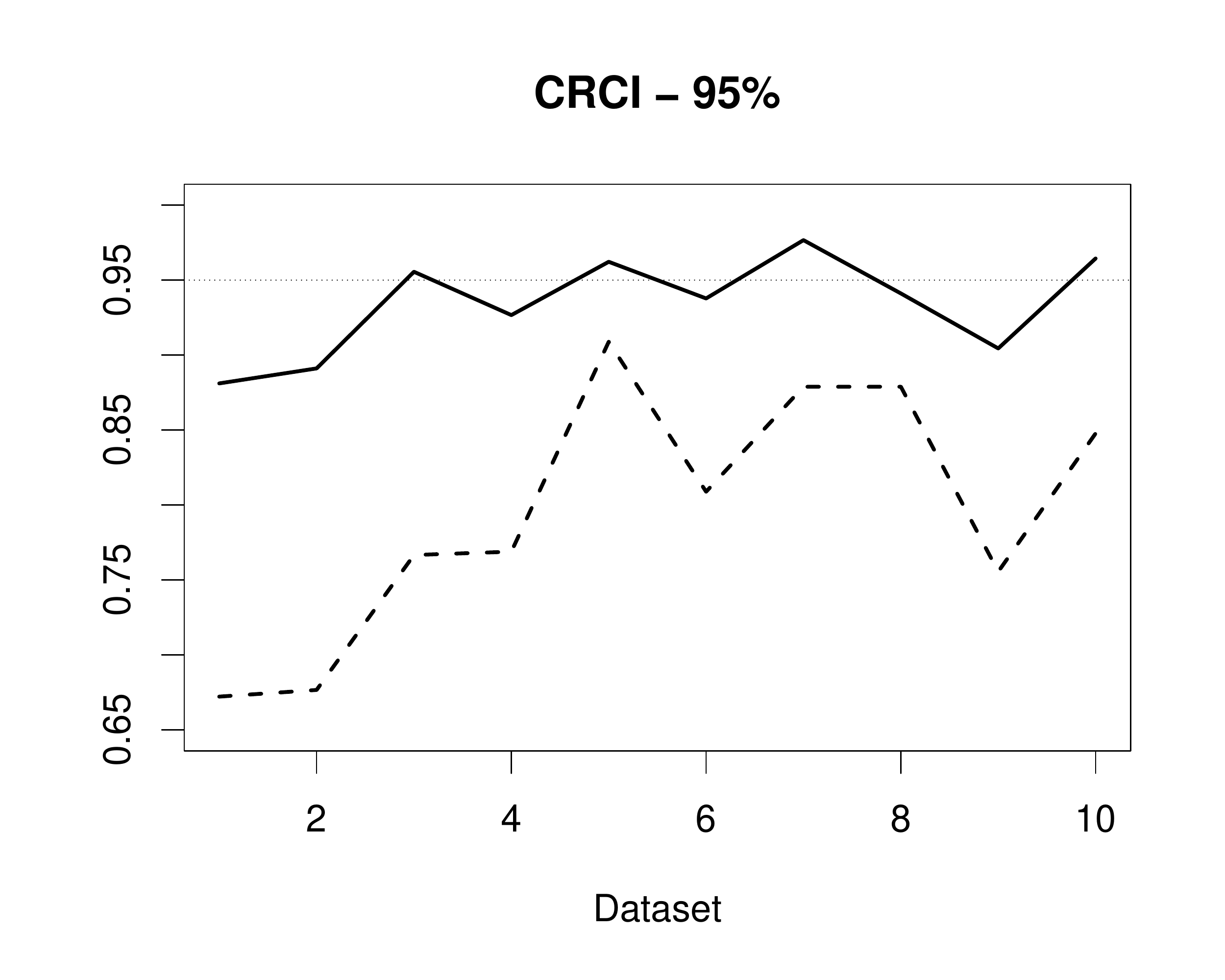}
	\caption{Quality measures of prediction for EPS (continuous line) and NPS (dashed line) models for simulated data with preferential sampling. \label{fig:simul:ps:predict_measures}}
\end{figure}

A regular grid of size $30 \times 30$ was considered and prediction was performed for both models.
Figure \ref{fig:simul:ps:predict_measures} shows the result and, as expected, the PS model
presented better predictive performance. Results for CRCI were obtained with $\alpha = 0.05$ but
similar results were obtained with credibility
intervals of 90\% and 99\% (graphs are presented in the supplementary material). Thus,
the EPS model proves to be the adequate route when preferentiality is present in the data.

\subsection{Data without preferential sampling}
It has just been shown that the exact model provided good results in the preferentiality context.
Now, the quality of the exact model is verified in a non preferential context. 10 datasets were
simulated under the traditional scenario of non-preferentiality and the sample size of each one
is presented in Table \ref{tab:simul:nps}. The same prior distributions of Section
\ref{sec:simul:ps} were adopted.
\begin{table}[b!]
	\centering \renewcommand\arraystretch{1.2}
	\begin{tabular}{lcccccccccc}
		\hline
		Dataset&1&2&3&4&5&6&7&8&9&10\\ \hline
		n&67&70&77&87&73&70&78&80&66&70\\
		\hline
	\end{tabular}
	\caption{Size of the simulated datasets without preferentiality.}\label{tab:simul:nps}
\end{table}

\begin{figure}[hbt!]
	\centering
	\includegraphics[scale=0.32]{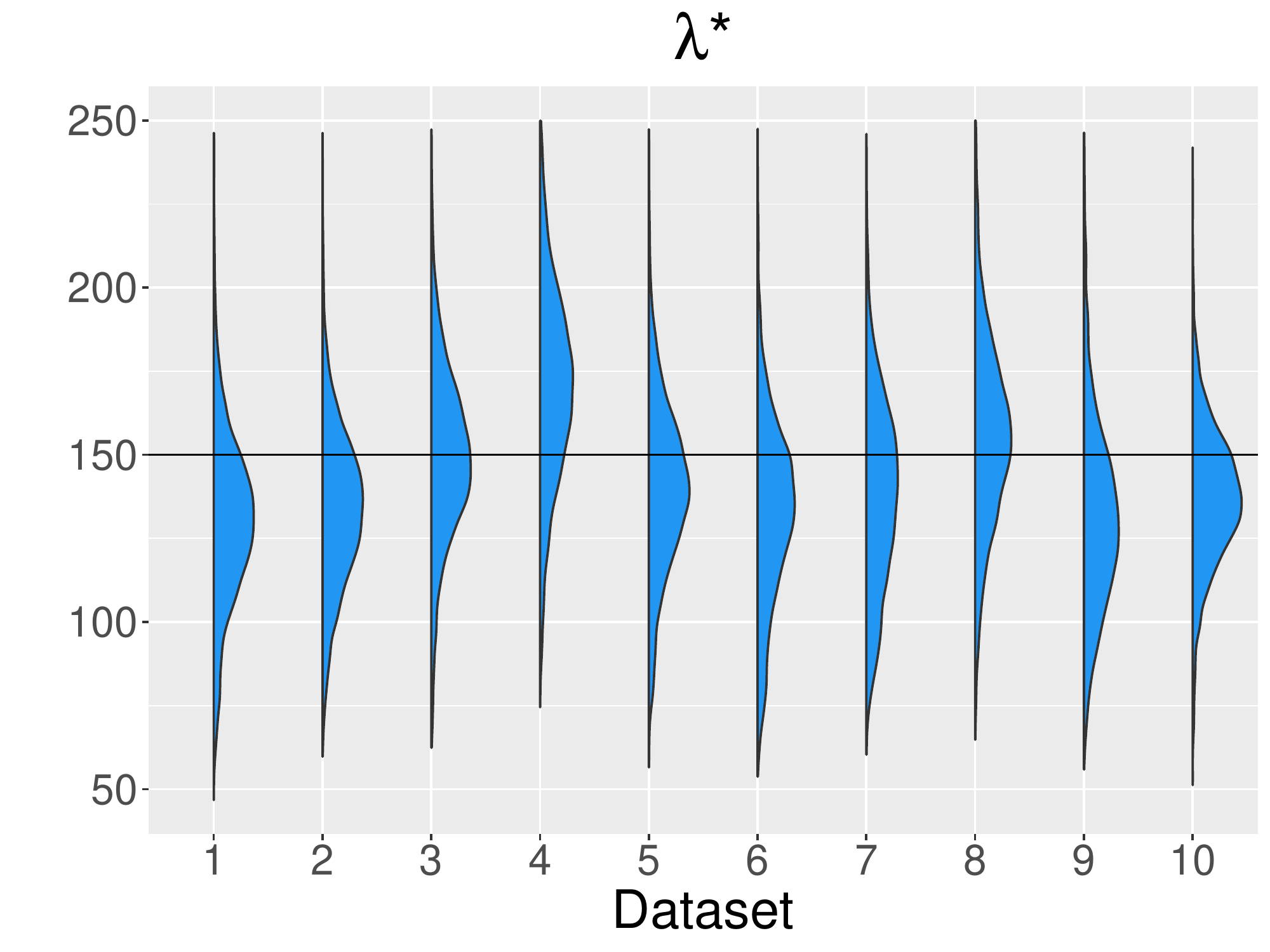}
	\includegraphics[scale=0.32]{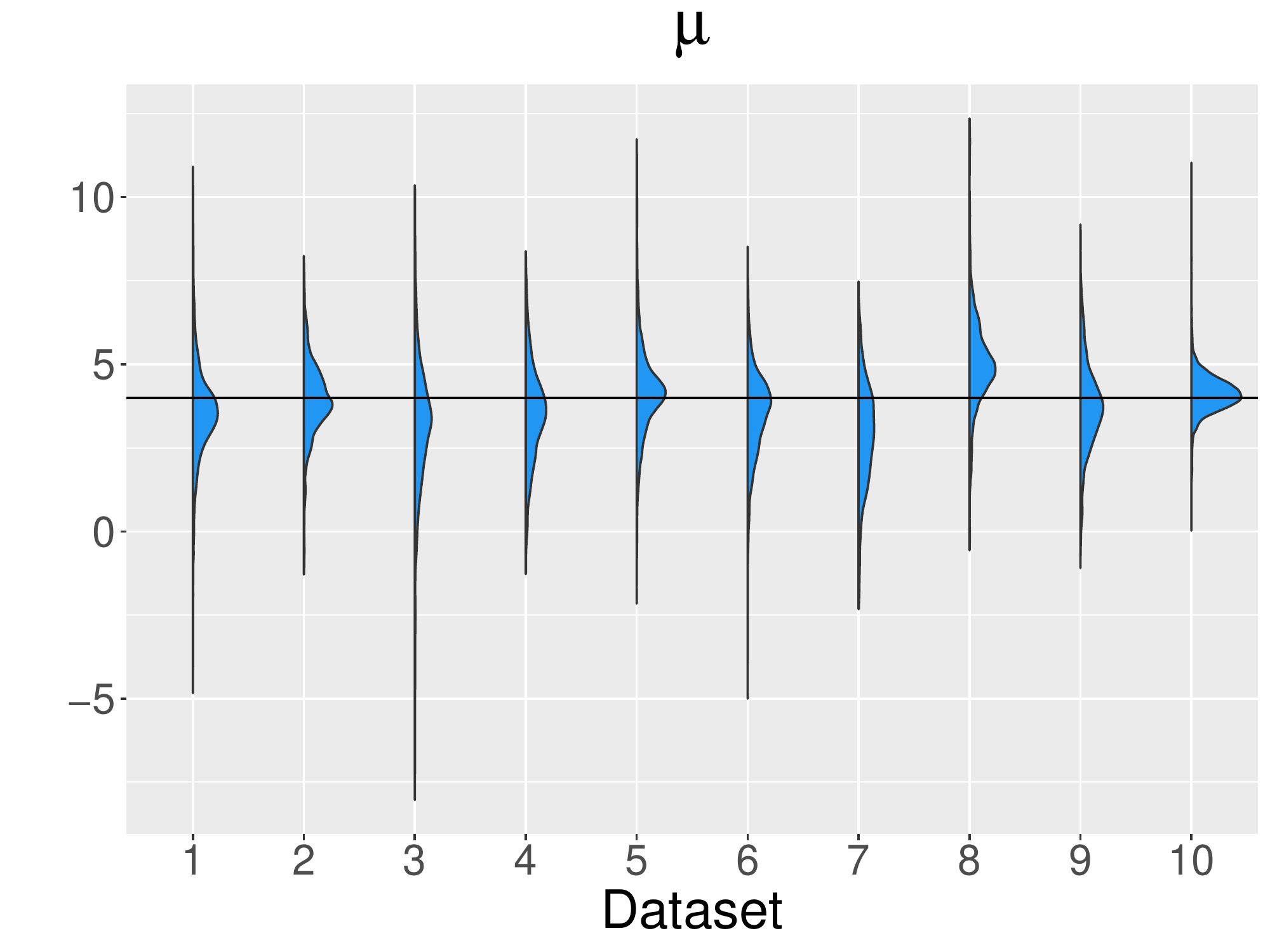}\\
	\includegraphics[scale=0.32]{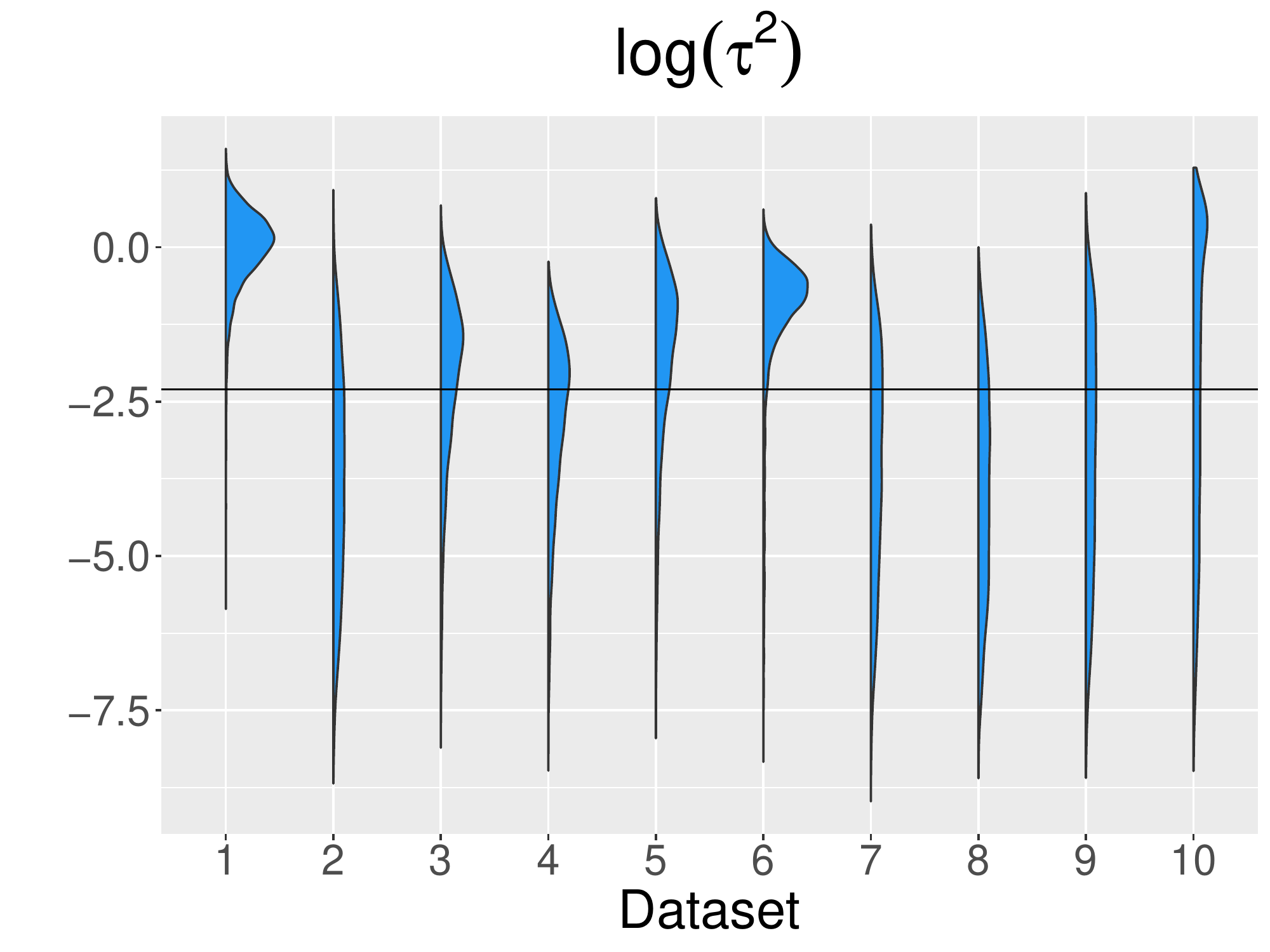}
	\includegraphics[scale=0.32]{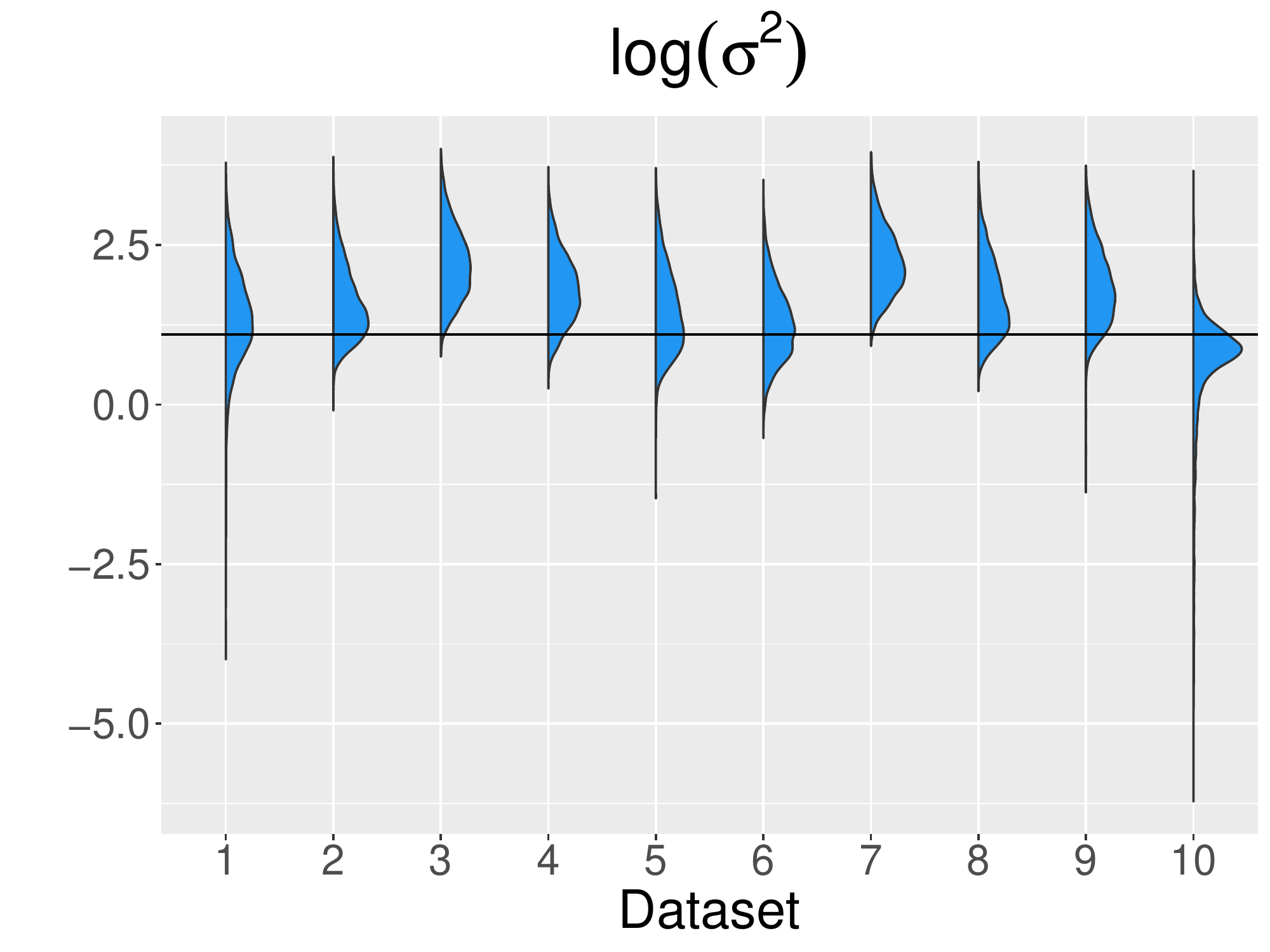}\\
	\includegraphics[scale=0.32]{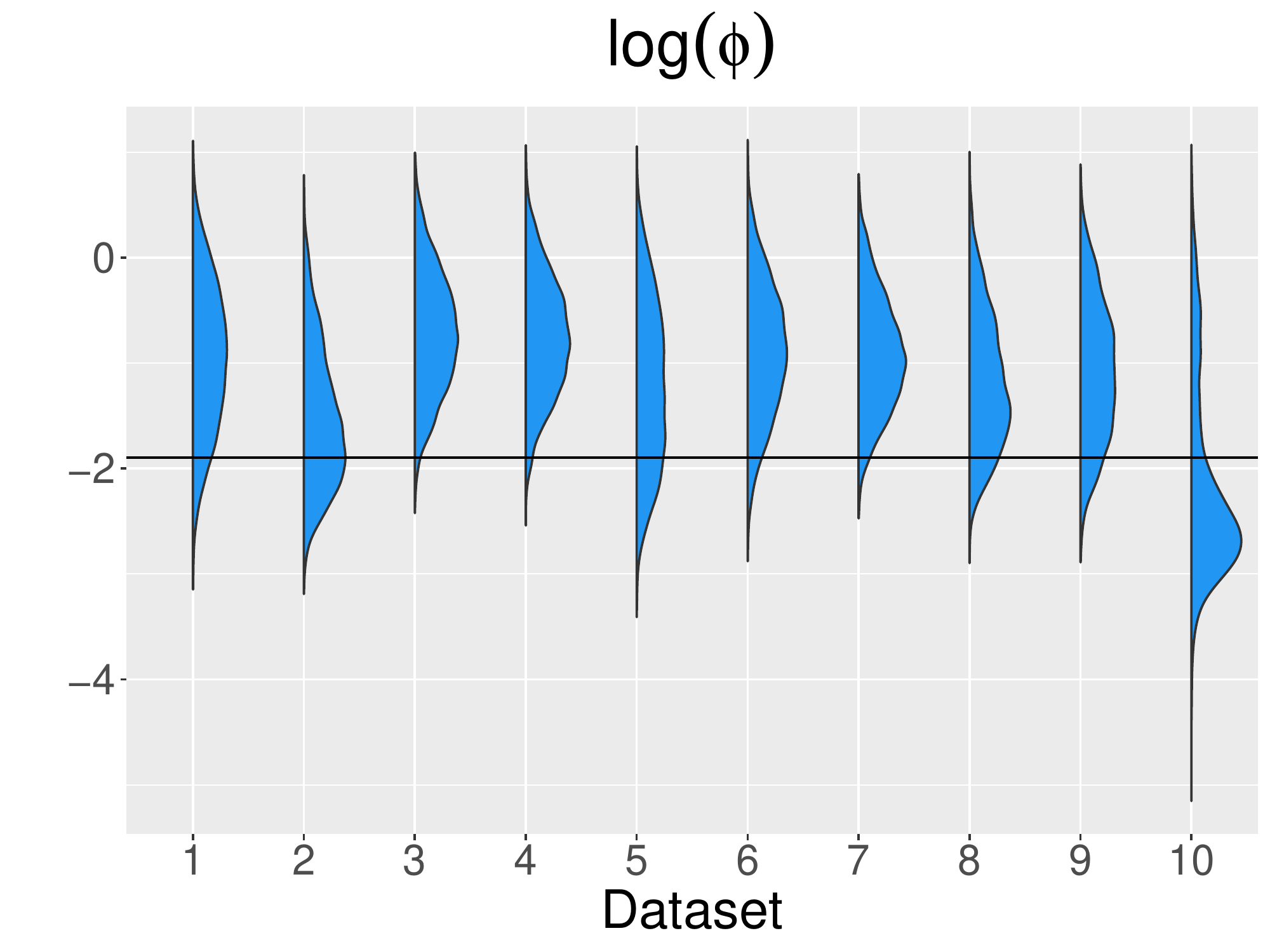}
	\includegraphics[scale=0.32]{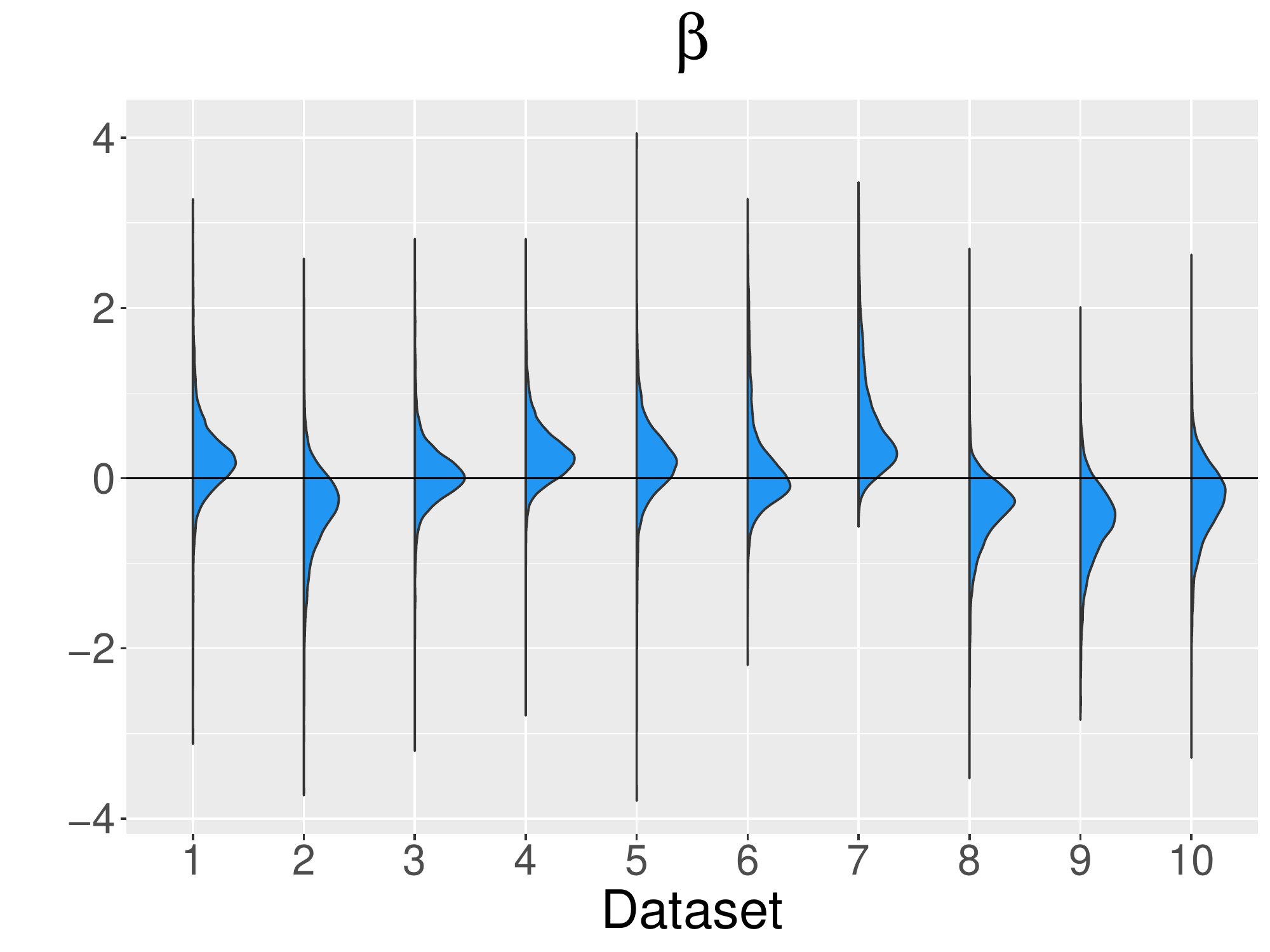}
	\caption{Posterior densities of the parameters of the EPS model for data simulated without preferential sampling. The horizontal line represents the true value of the parameter. \label{fig:simul:bayes:sec1:hist:10:beta0}}
\end{figure}

The posterior densities of the parameters estimated by the exact model are presented in Figure
\ref{fig:simul:bayes:sec1:hist:10:beta0}. As can be observed, good estimation was obtained
for all parameters. The posterior distribution of $\beta$ has most densities around the true value 0,
indicating that the model was able to capture the non-preferentiality of the data.

The quality of prediction was also compared for both models. It indicates that both models have
similar performance in predicting data at unobserved locations when there is no preferential
sampling (results are shown in the supplementary material).

These exercises suggest that little is lost by assuming preferentiality even when it is not present.
Estimation and prediction results do not seem to suffer in the comparison. These results differ
from those from the from previous section where wrongly assuming absence of preferentiality was
shown to deteriorate the performance of the wrong model. In both cases, they show a robust performance 
of the EPS model in a variety of different scenarios.

\subsection{Comparison of the exact and discrete models}

The performance of the commonly used discretized approach was also assessed.
A regular grid of size $15 \times 15$ over the unit square was considered and the discrete model
was constructed
as in \cite{ferreiragam} (the description of the model can be viewed in the supplementary material).
The datasets and prior distributions used here are the same of the Section \ref{sec:simul:ps}.

Similar estimates were observed for $\lambda^*$, $\mu$ and $\beta$ in terms of parameter estimation.
But better results for the variance and correlation parameters were obtained for the exact model
(see the posterior densities in the supplementary material). Since the exact approach does not use
approximations, the actual observed locations are used and simulations of points are made over the
continuous region, providing better results.

\begin{figure}[b!]
	\centering
	\includegraphics[scale=0.3,trim={2cm 1cm 1.5cm 1cm},clip]{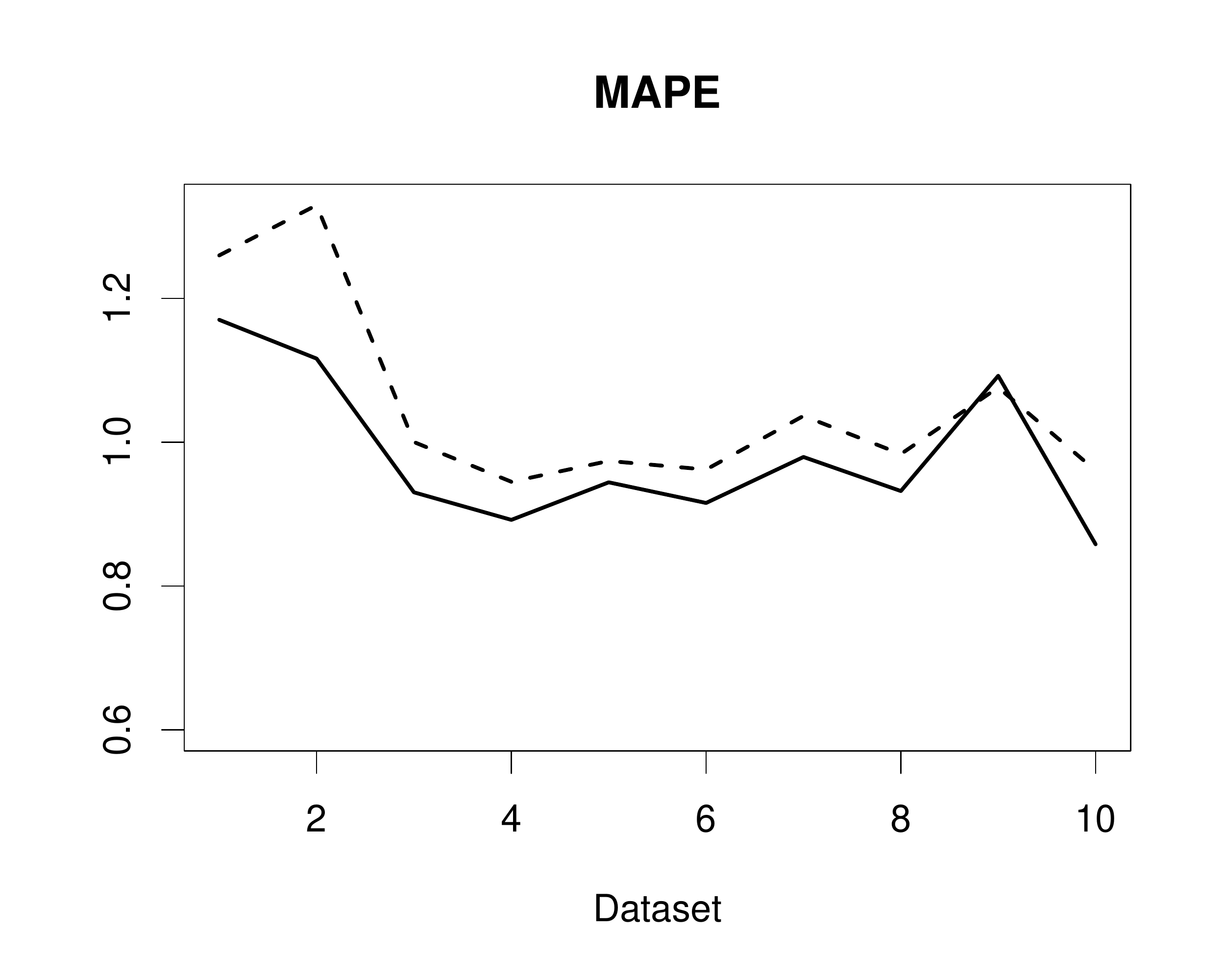}
	\includegraphics[scale=0.3,trim={2cm 1cm 1.5cm 1cm},clip]{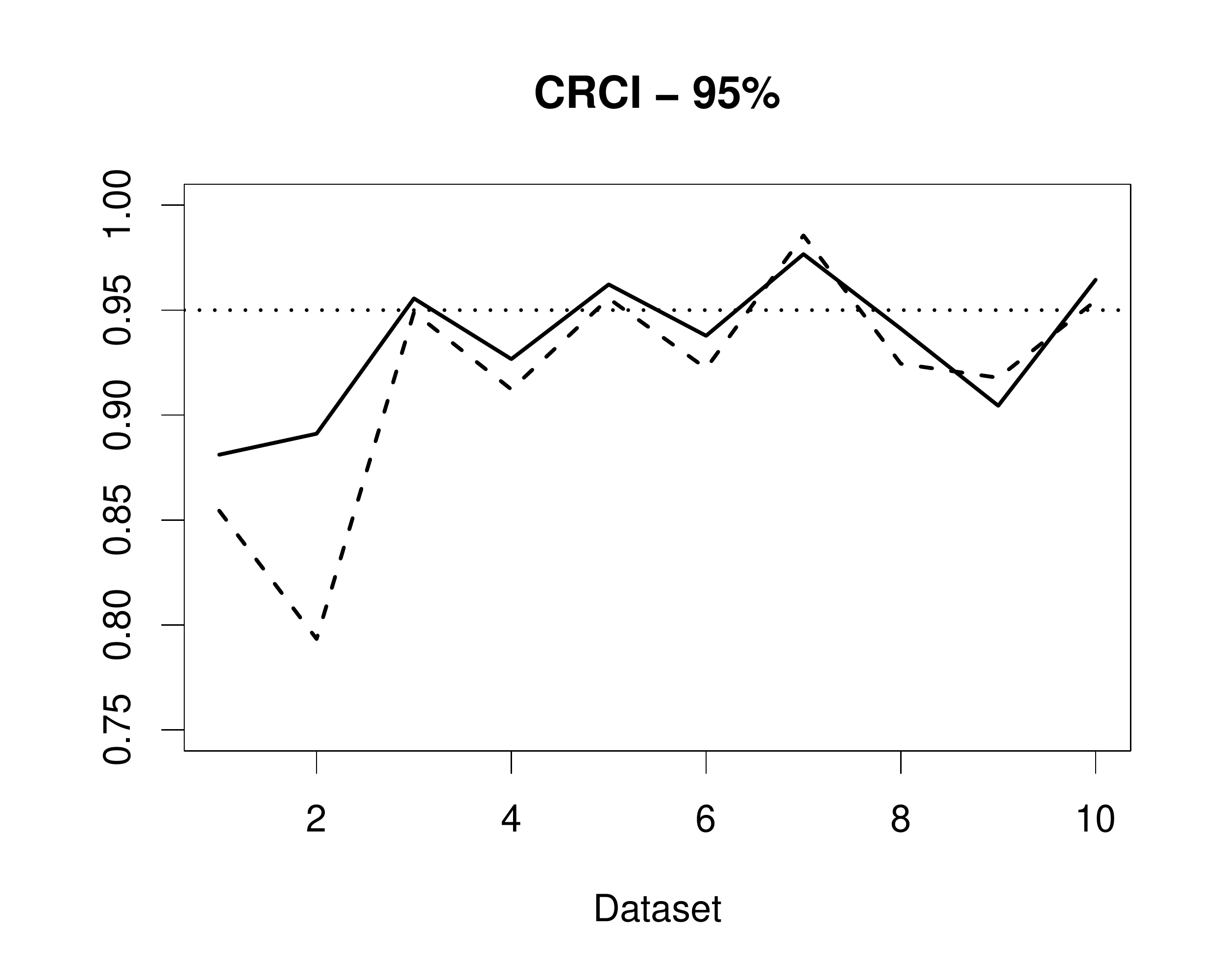}
	\caption{Graphs of prediction quality measures of the process $Y$ performed by the EPS (continuous line) and the DPS (dashed line) models considering the simulated datasets with preferential sampling. \label{fig:simul:bayes:sec1:pred:10:discreto_y}}
\end{figure}

The approaches were also compared in terms of prediction. Figure \ref{fig:simul:bayes:sec1:pred:10:discreto_y}
shows those measures calculated for both models. The exact model provided better prediction
measures in most of the datasets, which evidences the better performance of the exact methodology.
Similar results of CRCI considering 90\% and 99\% are presented in the supplementary material, and basically
confirm the results shown here.

\section{Application}\label{sec:application}

In this section, two real datasets are analyzed to illustrate the findings of the present work.
The first one is the radiation data in Germany, well known in the geostatistics literature as a dataset that does not present preferentiality \citep{dubois2005,ingram2007}. This data set is a good test for assessing the effectiveness of the exact preferential model in a context where no preferentiality was verified.
The second one is the Galicia moss data, another well known dataset in literature that presents preferentiality in
sampling \citep{fernandez2000,diggle2010,dinsdale2019}. This data is also analyzed with the discretized model, which enables the verification
of the possible benefits of the exact approach over the approximated discretization.

\subsection{Germany radiation data}

The Radioactivity Environmental Monitoring (REM) developed a data platform with the aim to make
data of radioactive monitoring available for researchers. The automatic mapping from updated data becomes useful
since its information updates occurs almost in real time. Then, the REM group
managed an web exercise, the Spatial Interpolation Comparison (SIC) of 2004, in which participants
could predict data using part of a real dataset. For further information, see \cite{Dubois2016}.
The data contain measures of mean rates of gamma radiation in Germany and consists in 1008
monitoring locations, in which only 200 chosen randomly location were provided. The competition
participants should perform prediction for the remaining 808 sites.

\begin{figure}[hbt!]
	\centering
	\subfigure{\includegraphics[scale=0.4,trim={0cm 0cm .5cm 0.5cm},clip]{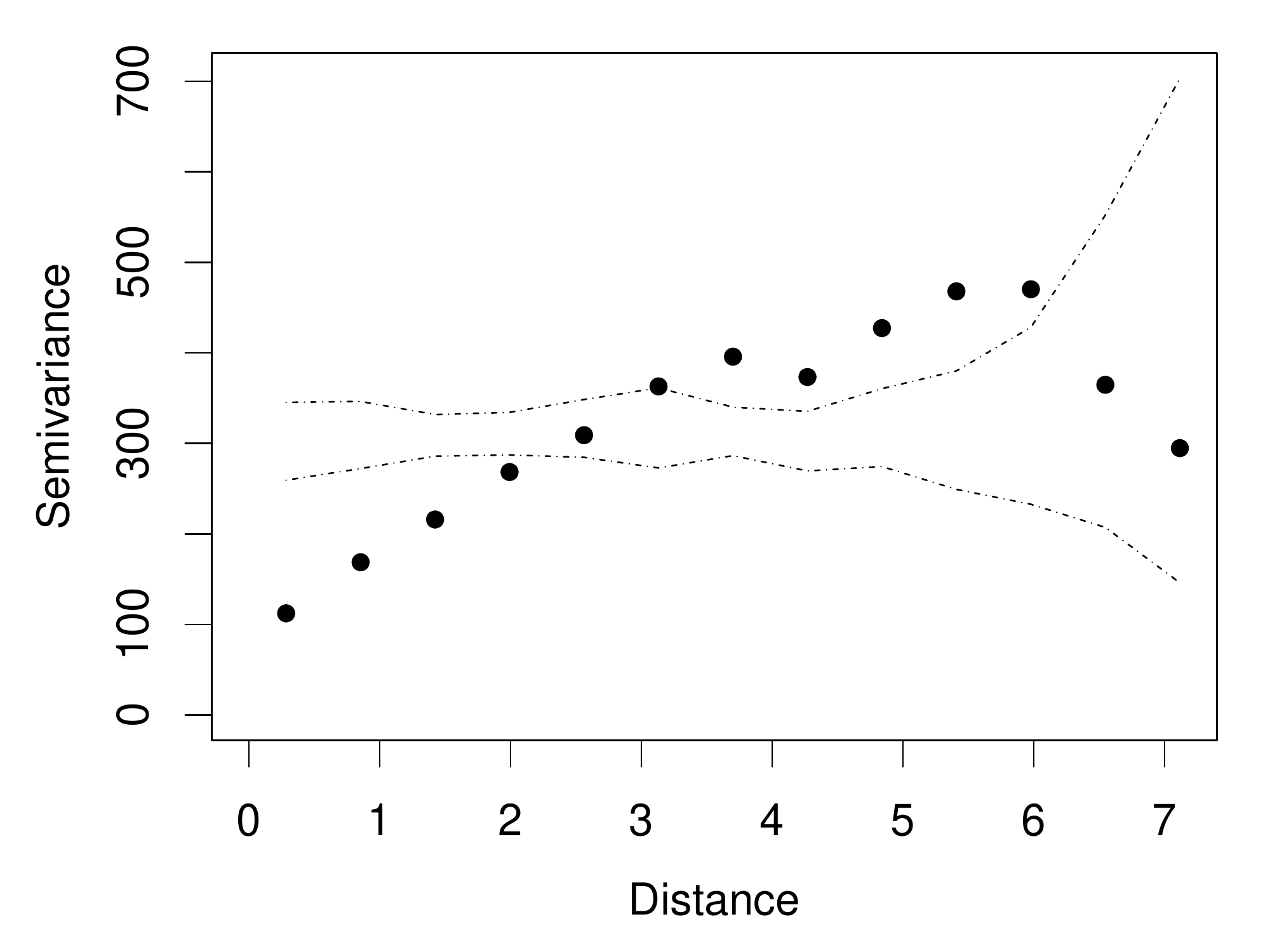}}
	\caption{Omnidirecional variogram and simulated envelope for radiation data in Germany. \label{fig:aplic:sic2004:variog4}}
\end{figure}

The variogram in Figure \ref{fig:aplic:sic2004:variog4} presents points out of the simulated
envelope (this was made by random permutations of data fixing its locations), indicating the
presence of spatial correlation. The exponential correlation function was considered and its range is equal to 3$\phi$ \citep{diggle2007}. Thus, through the empirical range (about 6) indicated in the variogram, the correlation parameter $\phi$ was fixed to be equal 2.

Data was analyzed with EPS and NPS models. The $\lambda^*$ parameter was truncated in $500/|B|$,
$B$ is the window considered of the region, indicating that the maximum mean number of points of
the $W$ process was set to 500. The following prior distributions were assumed: $\lambda^* \sim
Gamma(0.001,0.001)$, $\mu \sim N(0,10^6)$, $\tau^2 \sim IG(0.001,0.001)$, $\sigma^2 \sim
IG(0.001,0.001)$ and $\beta \sim N(0,1)$. The convergence of the Markov chains was verified through
graphical analysis (see supplementary material).

\begin{figure}[hbt!]
	\centering
	\includegraphics[scale=0.27]{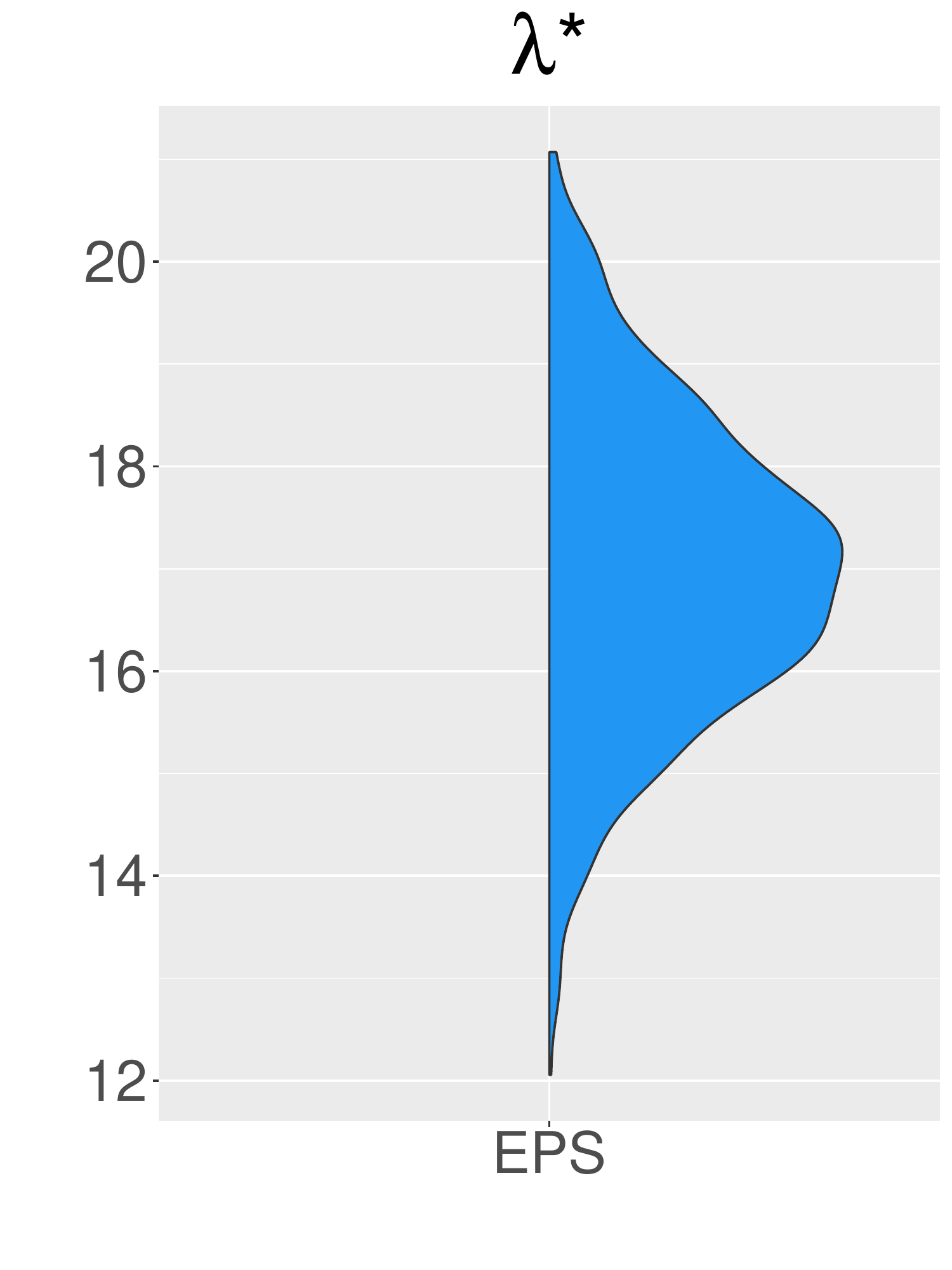}
	\includegraphics[scale=0.27]{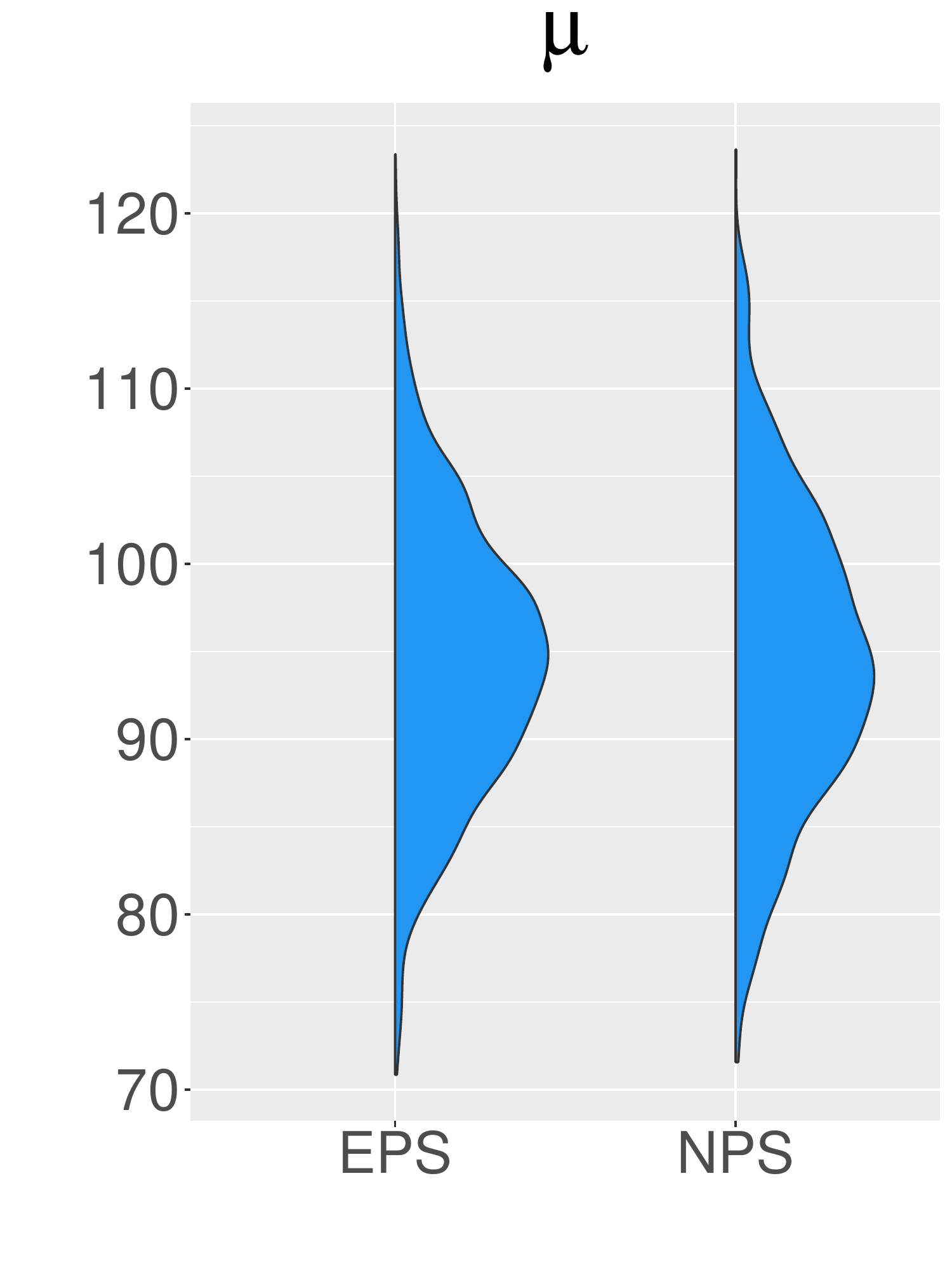}
	\includegraphics[scale=0.27]{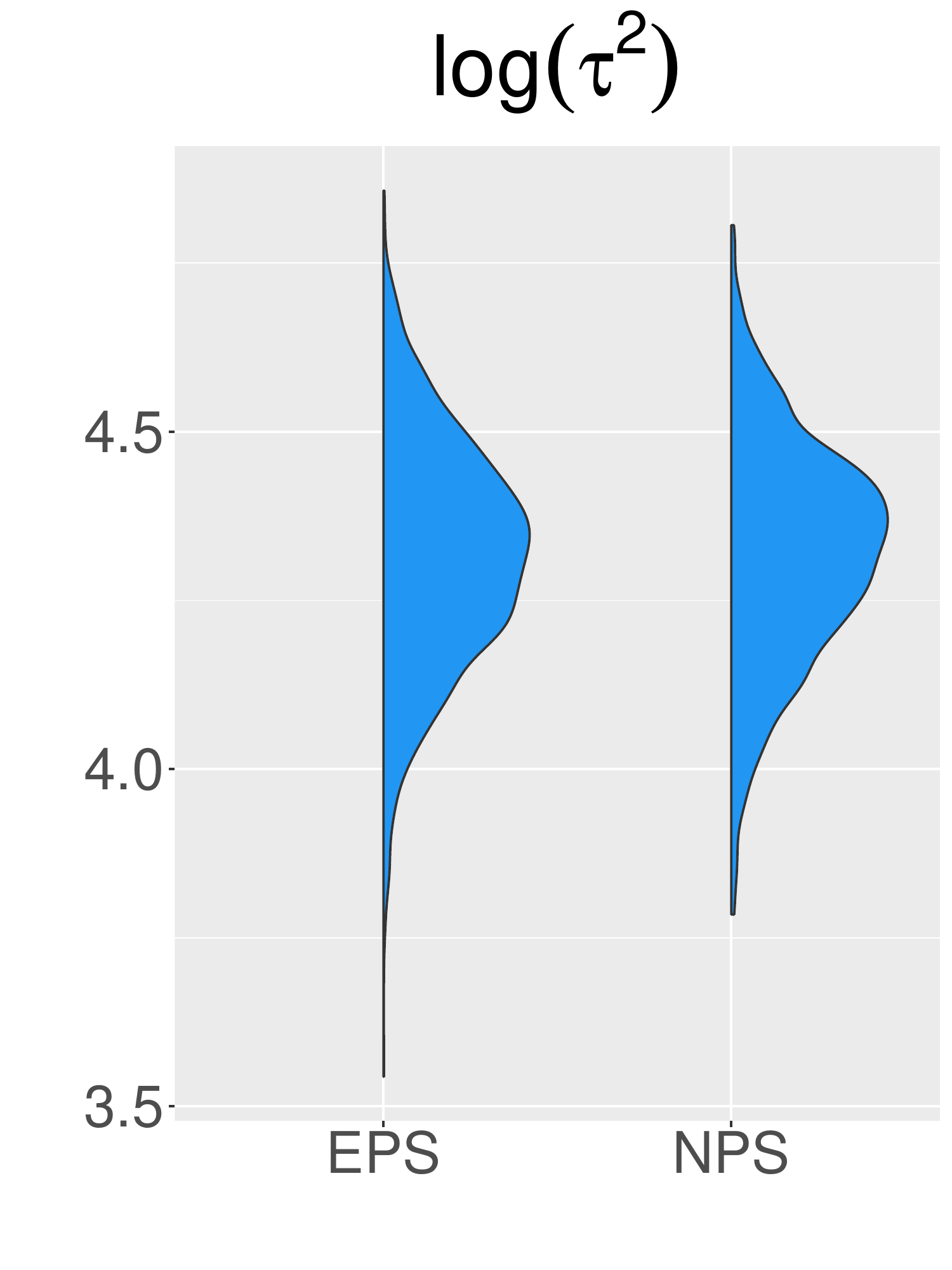}
	\includegraphics[scale=0.27]{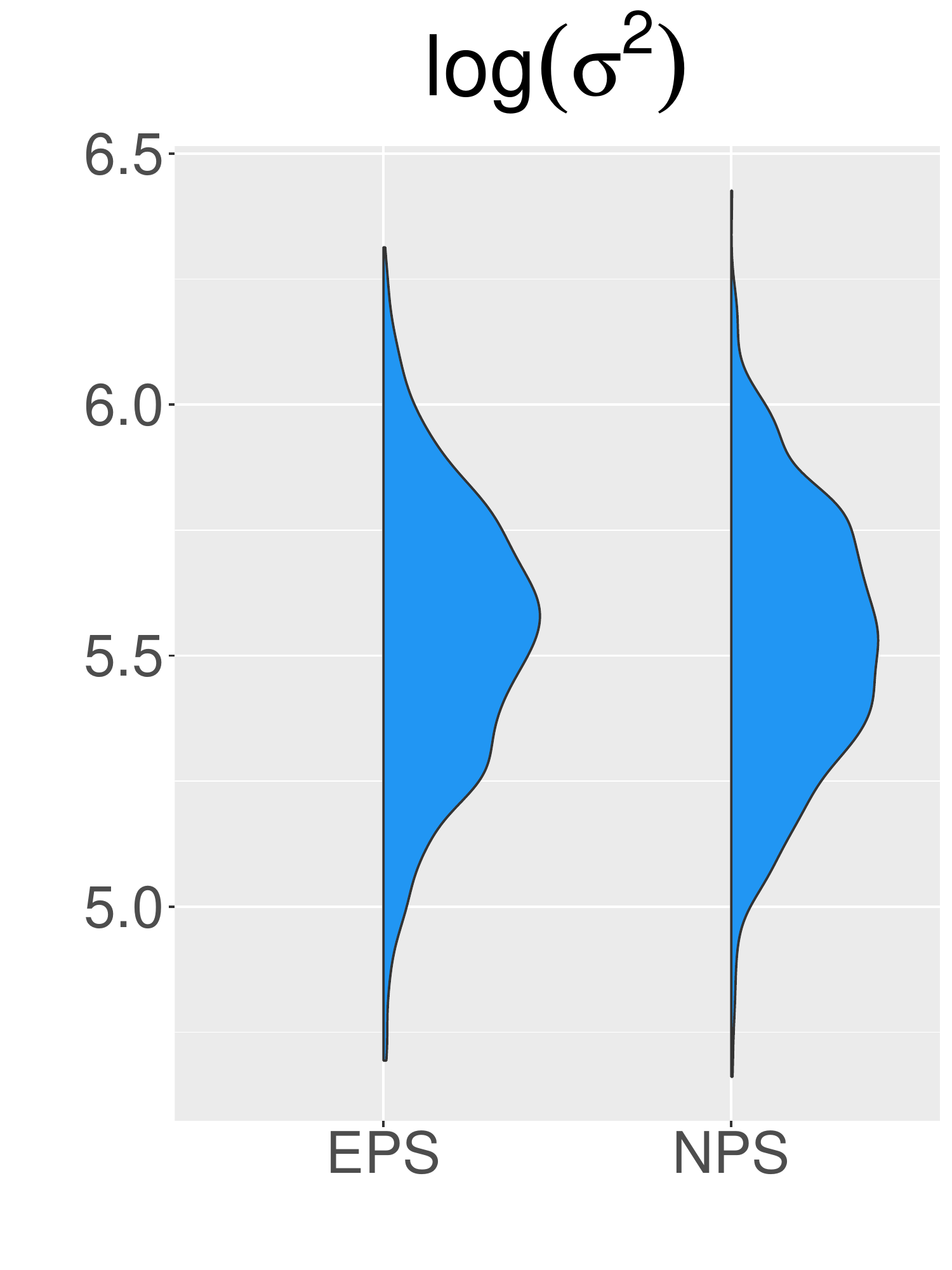}
	\includegraphics[scale=0.27]{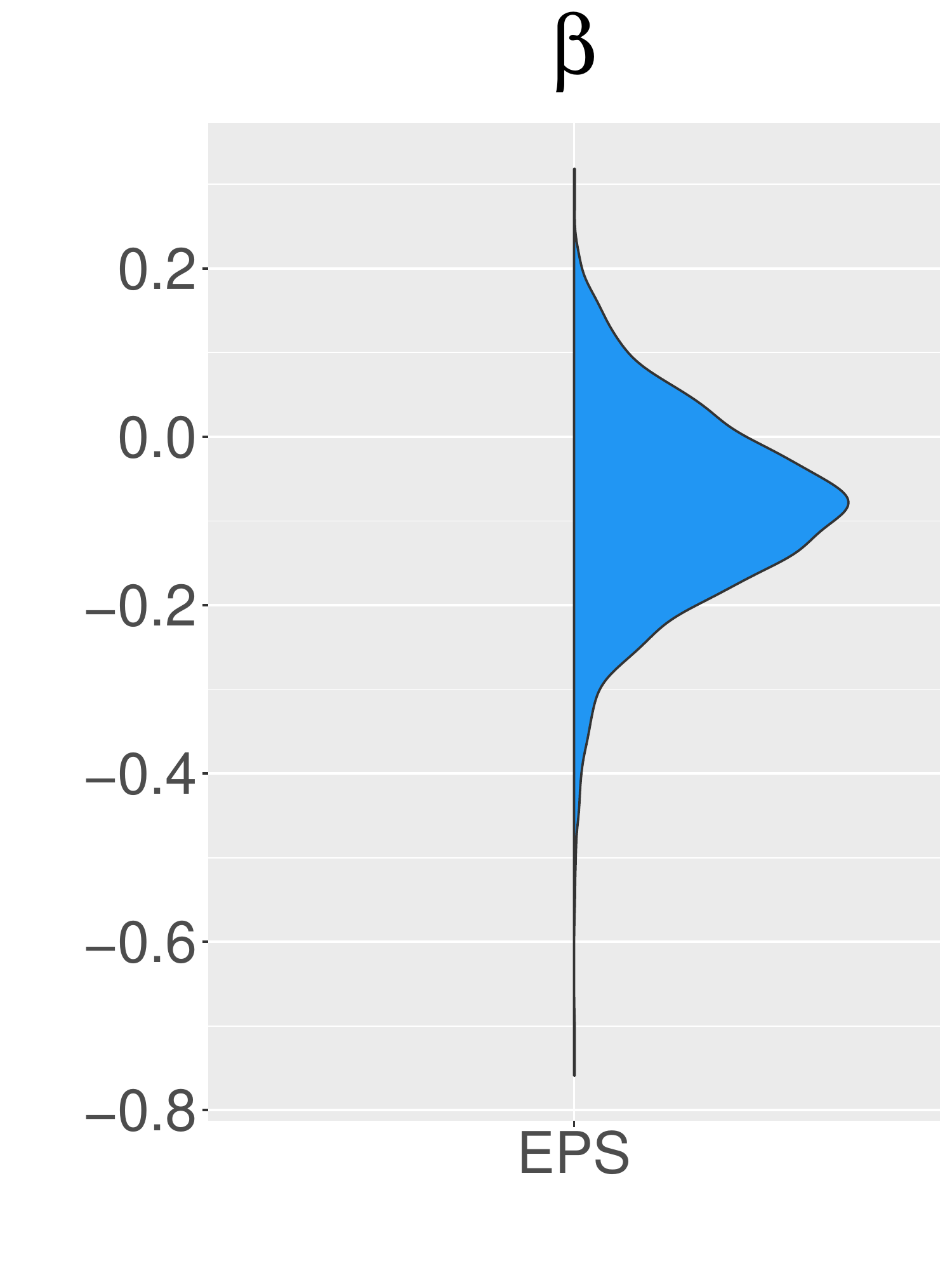}
	\caption{Posterior densities of model parameters for radiation data in Germany. \label{fig:aplic:sic2004:hist_}}
\end{figure}

Table \ref{tab:aplic:sic2004:infB:modelo_} shows summary measures of the posterior densities of
parameter for both models. The estimation of parameters that are common to both models is similar,
as expected. This is also confirmed with the $\beta$ estimation, with its estimated value near 0
and the 95\% credibility intervals containing 0. In Figure \ref{fig:aplic:sic2004:hist_}, all
posterior densities are unimodal and they are similar for both models considering the common parameters.
\begin{table}[hbt!]
	\centering \renewcommand\arraystretch{1.1}
	\begin{tabular}{ccrrrc}
		\hline
		Model&Parameter & Mean & Median & SE & CI(95\%) \\ \hline
		\multirow{5}{*}{EPS}&$\lambda^*$ &17.012 & 17.004 & 1.479 & [ 14.026 ; 20.015 ]\\
		&$\mu$ &94.670 & 94.571 & 7.795 & [ 80.165 ; 110.662 ]\\
		&$\tau^2$ &75.979 & 75.294 & 12.848 & [ 53.320 ; 104.075 ]\\
		&$\sigma^2$ &261.953 & 256.650 & 69.170 & [ 148.104 ; 418.897 ]\\
		&$\beta$ &-0.084 & -0.082 & 0.111 & [ -0.315 ; 0.130 ]\\ \hline
		\multirow{3}{*}{NPS}& $\mu$ &94.669 & 94.437 & 8.496 & [ 78.233 ; 112.714 ]\\
		&$\tau^2$ &76.194 & 75.960 & 12.378 & [ 53.181 ; 101.849 ]\\
		&$\sigma^2$ &259.590 & 251.229 & 67.085 & [ 155.017 ; 403.486 ]\\
		\hline
	\end{tabular}
	\caption{Summary measures of the posterior density distributions of the parameters of the EPS and NPS  models for radiation data in Germany.}\label{tab:aplic:sic2004:infB:modelo_}
\end{table}		

\begin{figure}[hbt!]
	\centering
	\subfigure{\includegraphics[scale=0.3]{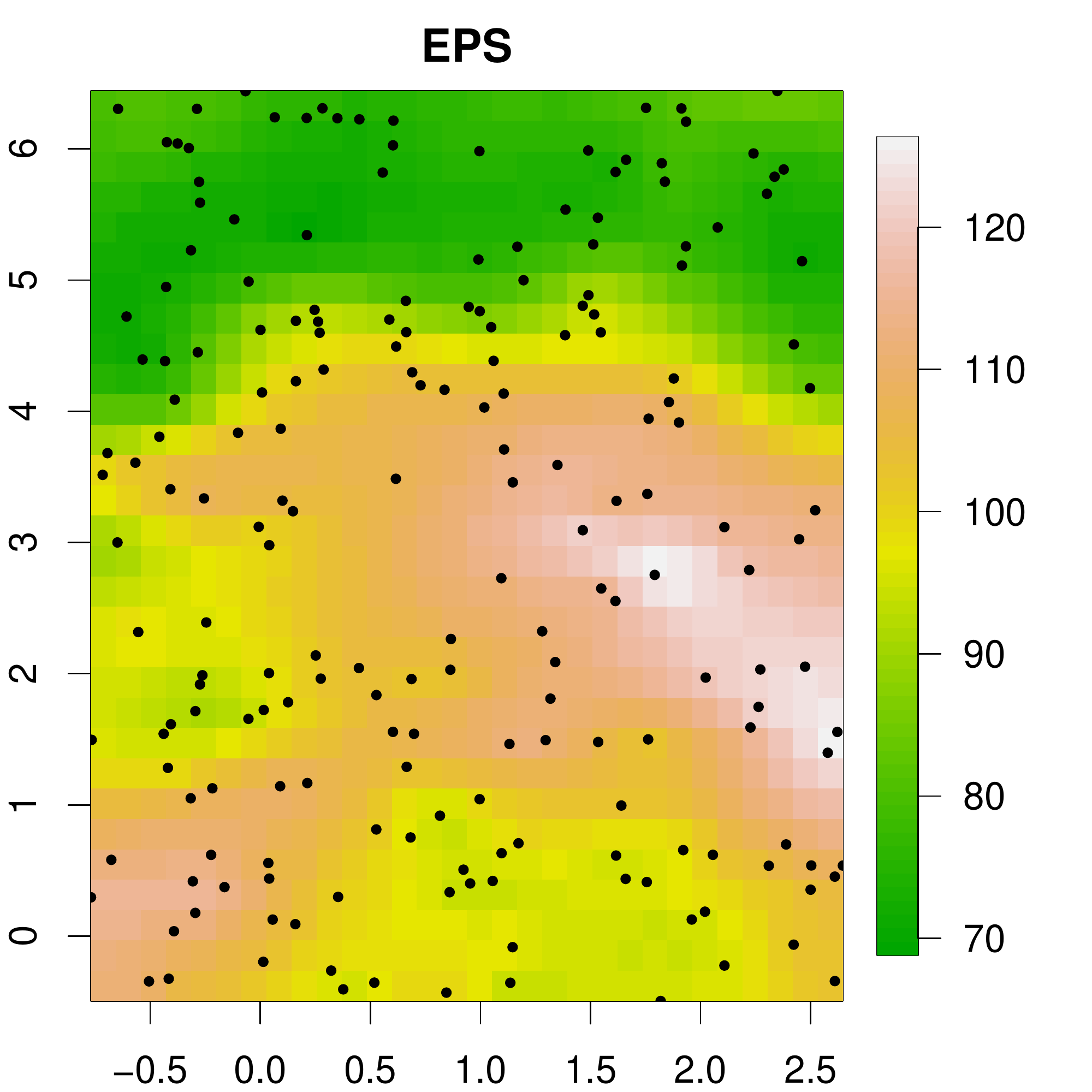}
		\includegraphics[scale=0.3]{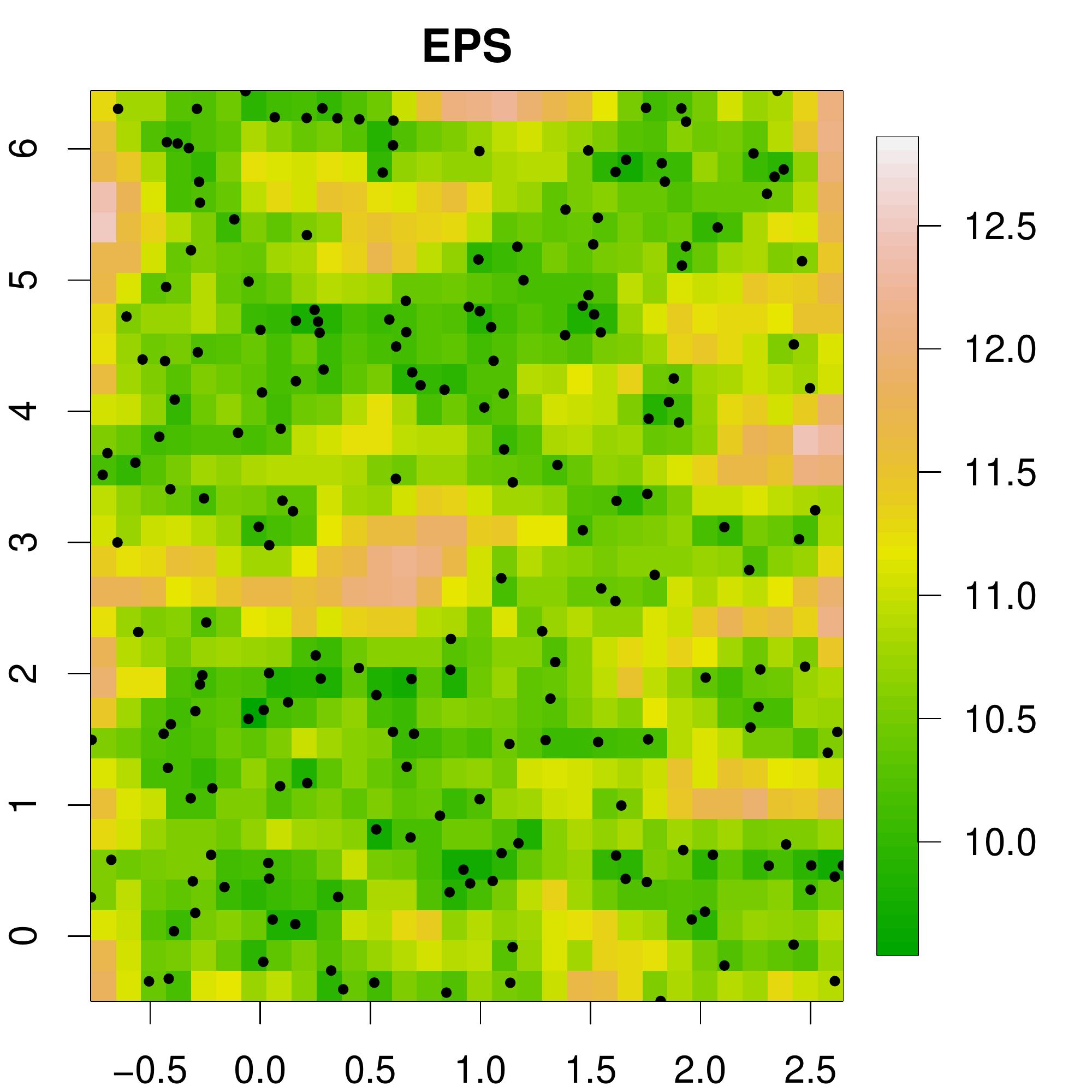}}\\
	\subfigure{\includegraphics[scale=0.3]{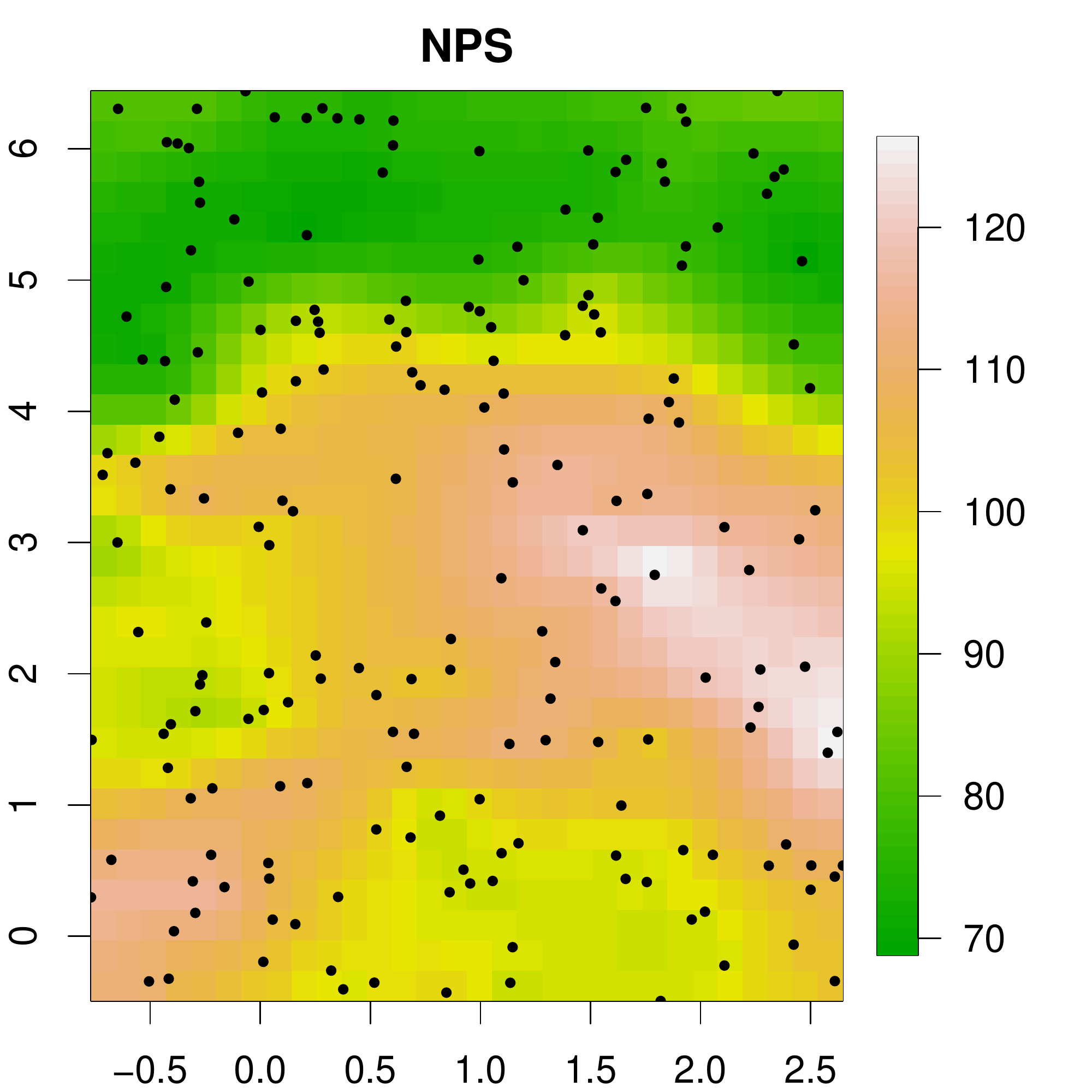}
		\includegraphics[scale=0.3]{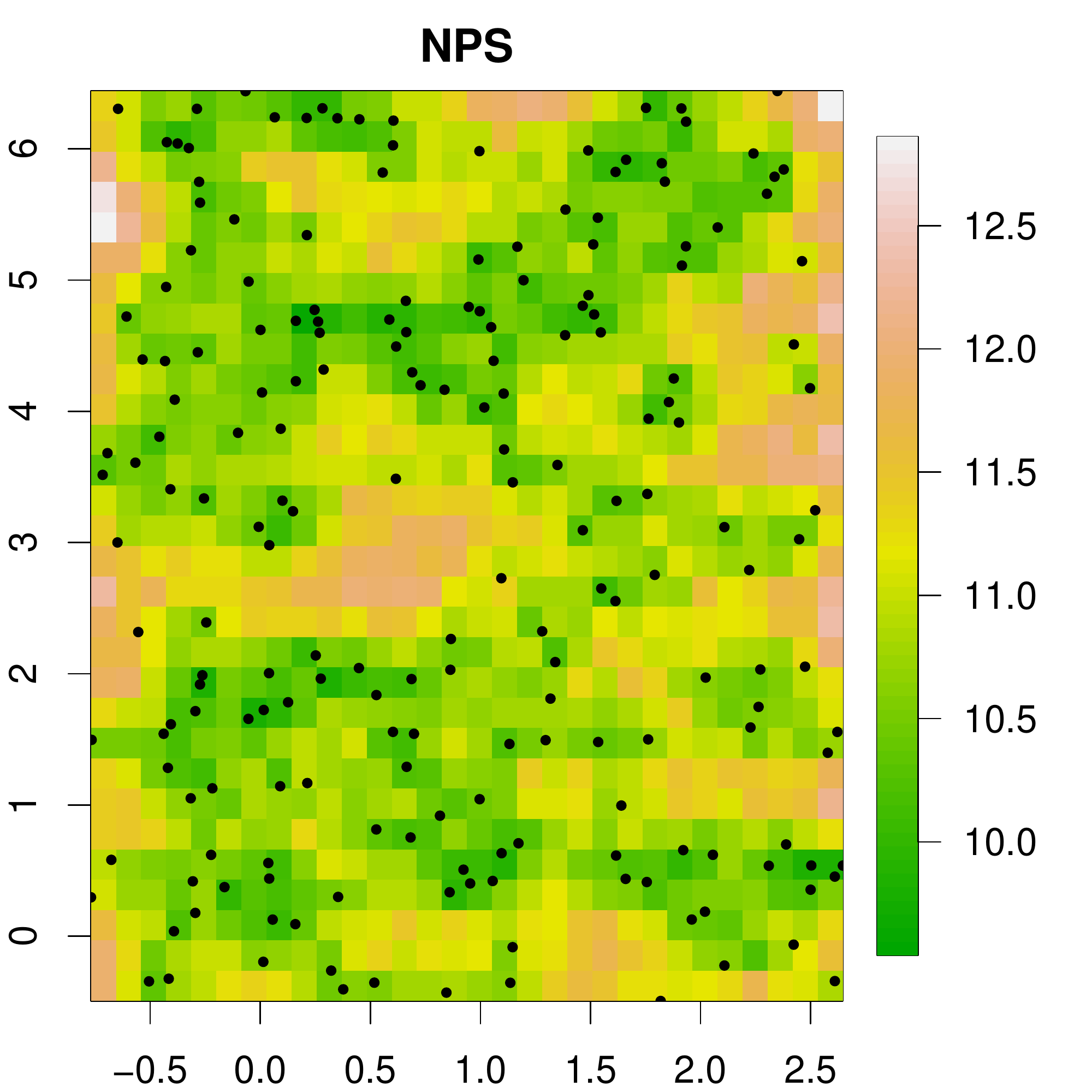}}
	\caption{Predicted maps of $Y$ (first column) and predicted errors (second column) for radiation data in Germany. \label{fig:aplic:sic2004:pred_mapa}}
\end{figure}

The process $Y$ was predicted at the 808 remaining sites of the dataset and MAPE was calculated.
The MAPE values obtained for the EPS and NPS model were quite similar with respective values 9.073 and 9.075.
The NPS model is more adequate since data does not shows preferentiality.
But, the results indicates that the EPS model performs as well as the NPS model in this context.
These results corroborate the findings from the previous section, which indicate that there is no
substantial loss in performance by using the EPS model for non preferential data. Figure
\ref{fig:aplic:sic2004:pred_mapa} shows the predicted maps of data over a $30 \times 30$ regular
grid and similar maps are observed between the models, with lower predicted errors corresponding to
the EPS model. Thus, the exact model showed good performance even in the non preferential sampling
context with a real dataset scenario.

\subsection{Galicia moss data}

\begin{figure}[htb!]
	\centering
	\subfigure{\includegraphics[scale=0.35,trim={0cm 0cm 1.5cm 2.5cm},clip]{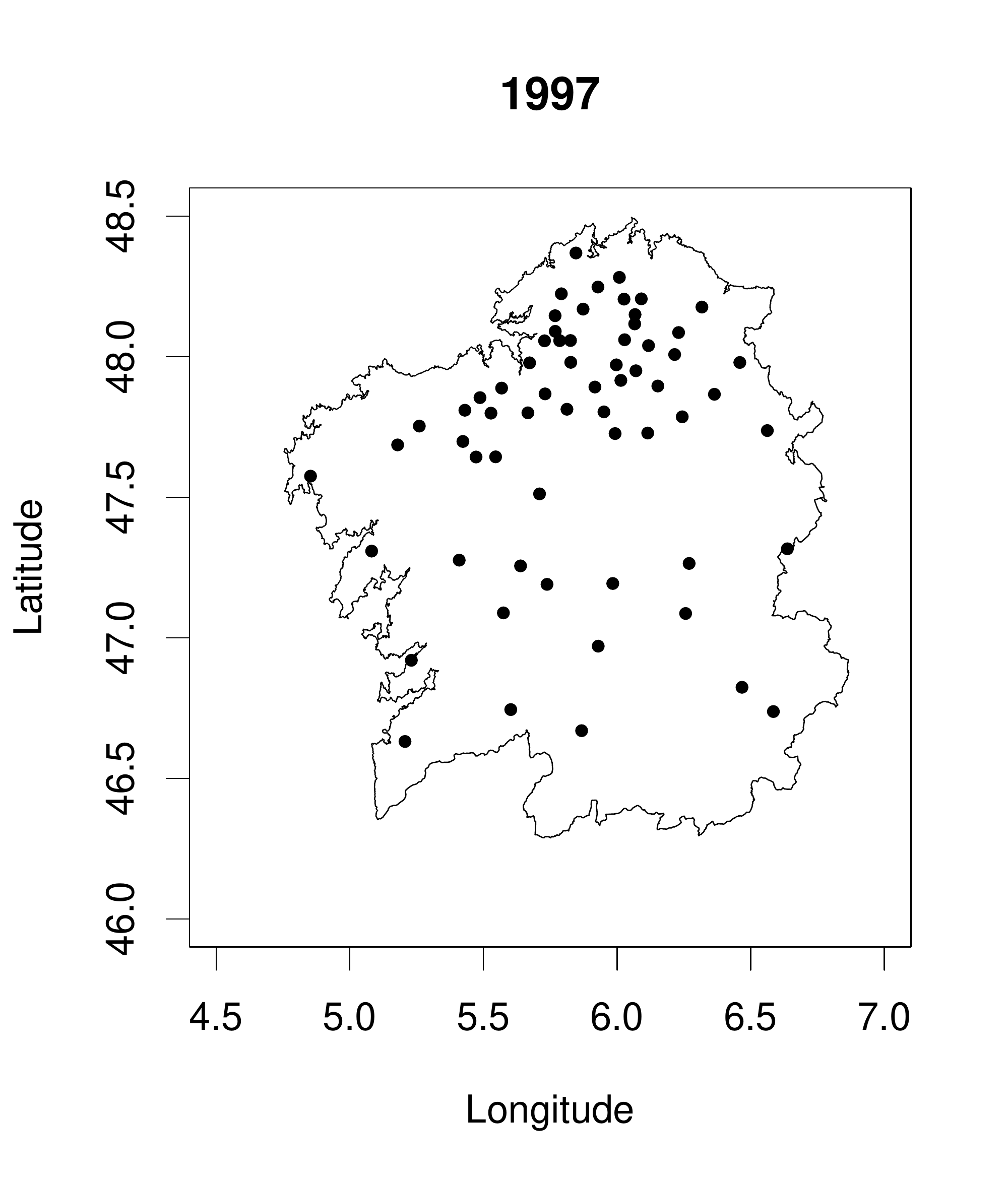}}
	\caption{Sampling locations of Galicia moss data. \label{fig:aplic:moss:mapa}}
\end{figure}

The data consists of measures of lead concentrations in samples of moss in
Galicia, northern Spain. The uptake of heavy metals in mosses occurs mainly from atmospheric
deposition, which turns mosses into biomonitoring of pollution. The study was conducted over a number of years
but here we concentraqted on the analysis of the data collected in
October 1997. Figure \ref{fig:aplic:moss:mapa} shows the sampling design. The choice of
sampling locations was conducted more intensively in subregions where high lead concentration
was expected, turning this sampling design potentially preferential. As a result, this data set is frequently used as an example of sampling preferentiality in the geostatistics literature.

Therefore, this data was analyzed only with the models that contemplate preferentiality EPS and DPS
in order to assess the possible effect of taking an exact stand at preferentiality.
A regular grid of $15\times 15$ was
considered for the DPS model. The following prior distributions were adopted: $\lambda^* \sim
Gamma(0.001,0.001)$, $\mu \sim N(0,10^6)$, $\tau^2 \sim IG(0.001,0.001)$, $\sigma^2 \sim
IG(0.001,0.001)$, $\phi \sim Gamma(2,4)$, and $\beta \sim N(0,1)$. The convergence of the Markov
chains was verified through graphical analysis (see supplementary material).

\begin{figure}[hbt!]
	\centering
	\includegraphics[scale=0.27]{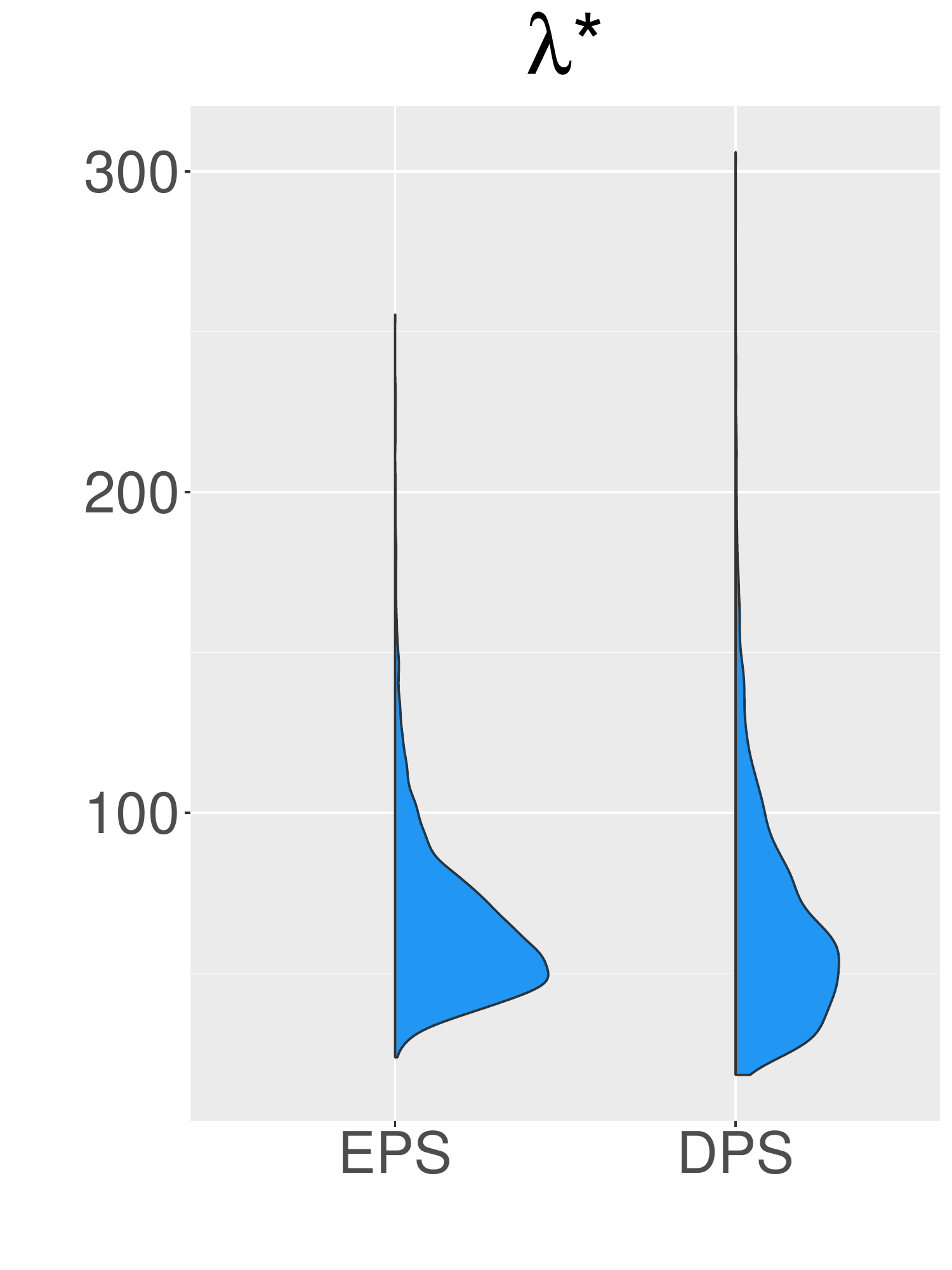}
	\includegraphics[scale=0.27]{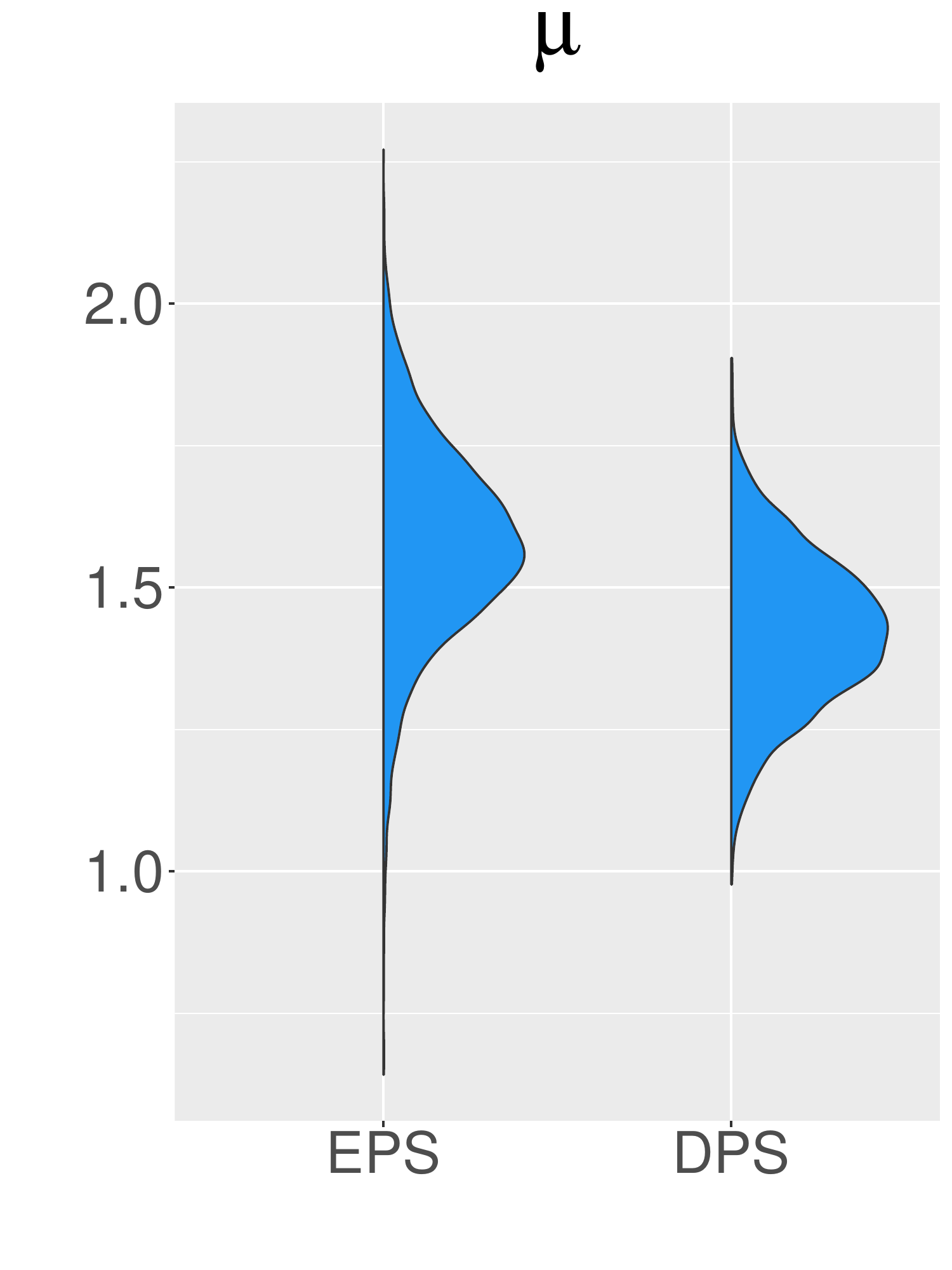}
	\includegraphics[scale=0.27]{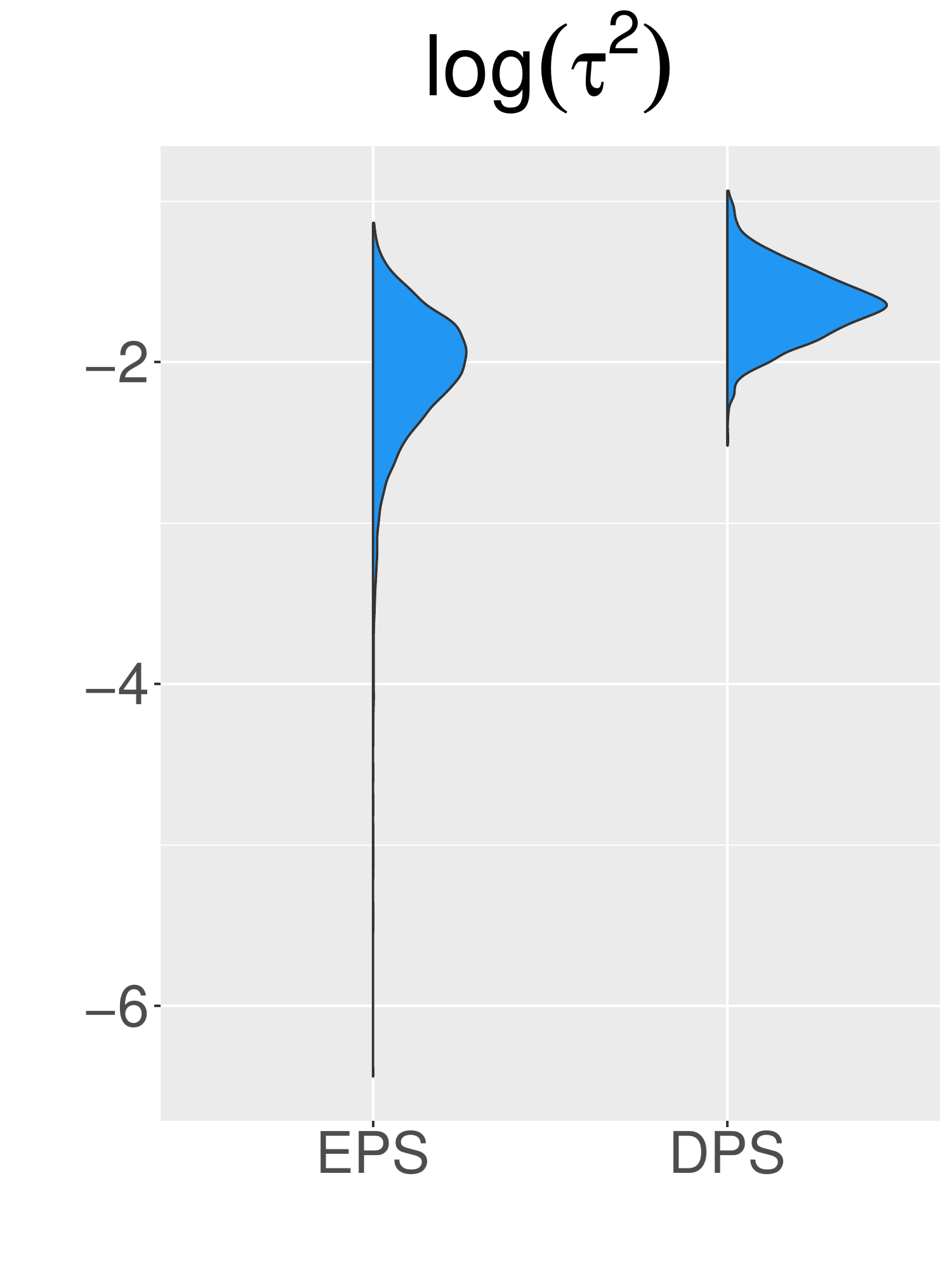}\\
	\includegraphics[scale=0.27]{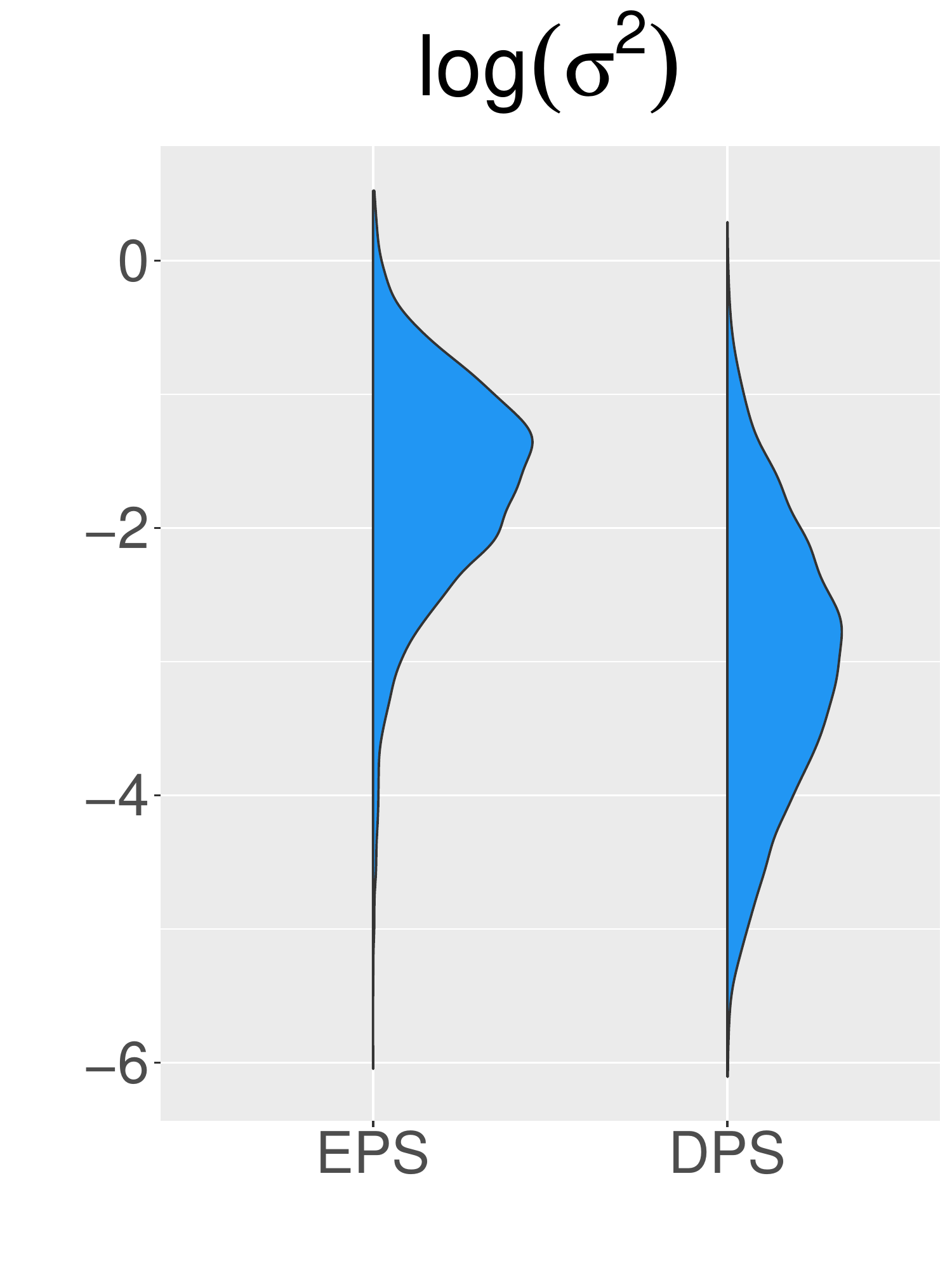}
	\includegraphics[scale=0.27]{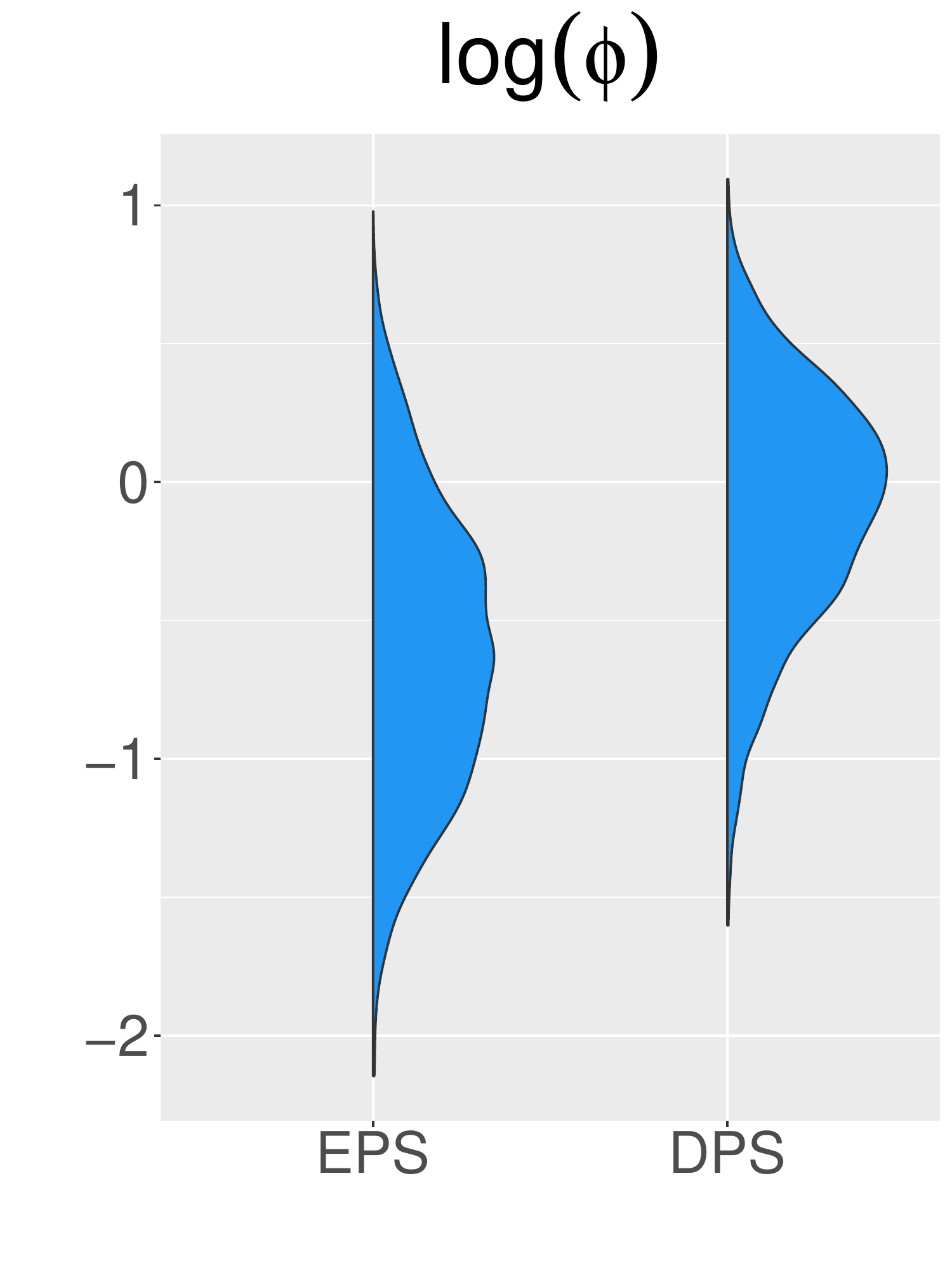}
	\includegraphics[scale=0.27]{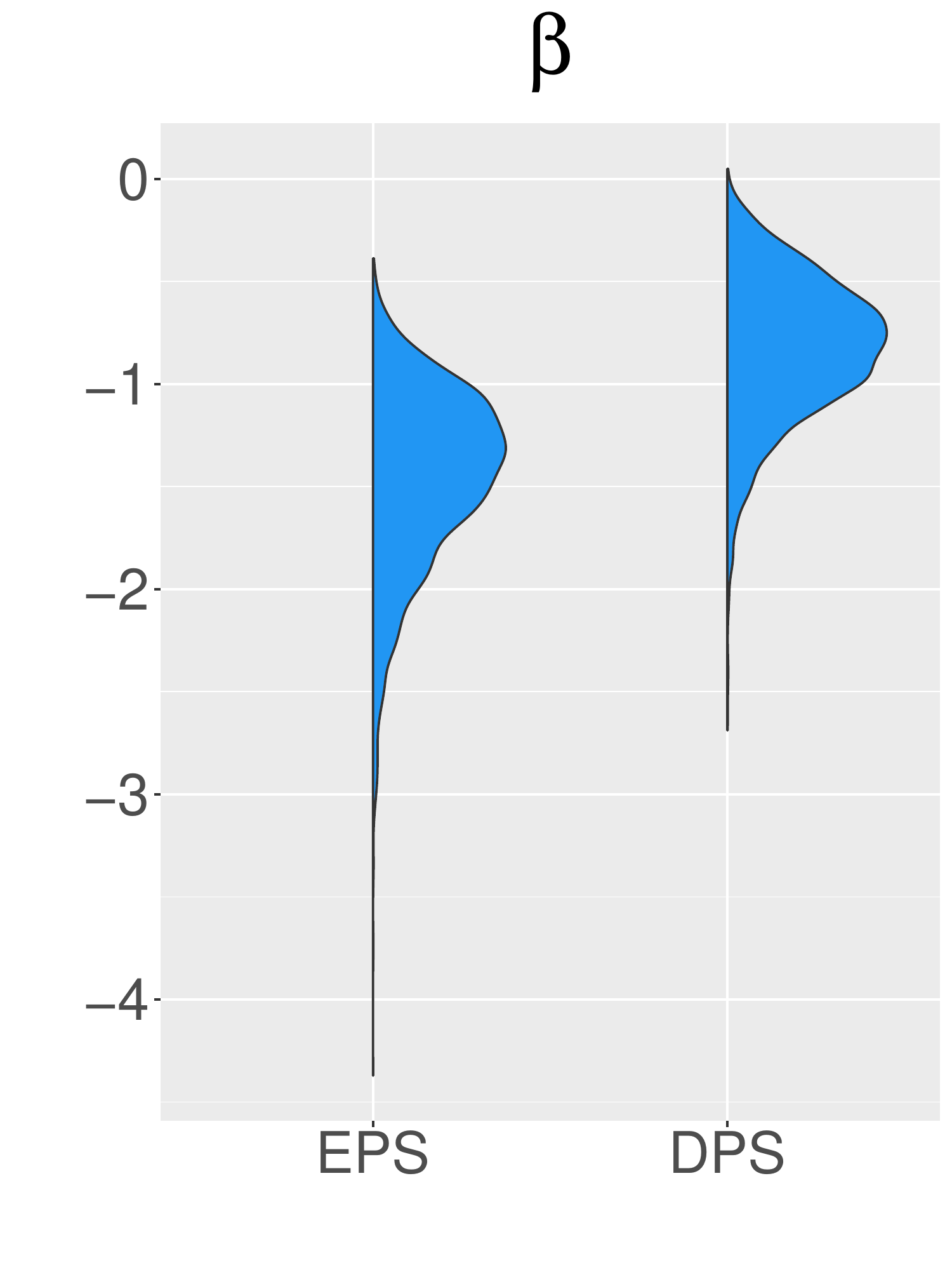}
	\caption{Posterior densities of model parameters for Galicia moss data in 1997.
		\label{fig:aplic:moss:hist_1997}}
\end{figure}

Table \ref{tab:aplic:galicia:infB:modelo_} shows summary measures and Figure
\ref{fig:aplic:moss:hist_1997} presents the posterior densities of parameter for the models.
Estimation of $\beta$ in both models indicated large concentration of posterior probability over negative values and
basically not including 0. These results strongly indicate that there could be preferentiality in this survey, in line with previous analyses of this dataset.
The estimation with the DPS model indicate higher values of the nugget effect and lower values for
$\sigma^2$. The effect of the discretization can also be seen in the estimation
of $\beta$. It is clearly different from 0 for both the exact and the discretized models,
but with dampened values for the latter in comparison to the values obtained with the exact PS model.
\begin{table}[hbt!]
	\centering \renewcommand\arraystretch{1.1}
	\begin{tabular}{ccrrrc}
		\hline
		Model&Parameter & Mean & Median & SE & CI(95\%) \\ \hline
		\multirow{5}{*}{EPS}&$\lambda^*$ &64.843 & 59.330 & 24.640 & [ 34.449 ; 128.702 ]\\
		&$\mu$ &1.576 & 1.580 & 0.177 & [ 1.196 ; 1.917 ]\\
		&$\tau^2$ &0.135 & 0.133 & 0.048 & [ 0.046 ; 0.237 ]\\
		&$\sigma^2$ &0.245 & 0.203 & 0.184 & [ 0.031 ; 0.686 ]\\
		&$\phi$ &0.615 & 0.533 & 0.332 & [ 0.212 ; 1.486 ]\\
		&$\beta$ &-1.458 & -1.394 & 0.441 & [ -2.495 ; -0.763 ]\\ \hline
		\multirow{5}{*}{DPS}&$\lambda^*$ &65.520 & 58.210 & 33.395 & [ 25.093 ; 150.663 ]\\
		&$\mu$ &1.423 & 1.423 & 0.132 & [ 1.159 ; 1.677 ]\\
		&$\tau^2$ &0.198 & 0.193 & 0.044 & [ 0.128 ; 0.298 ]\\
		&$\sigma^2$ &0.081 & 0.052 & 0.093 & [ 0.007 ; 0.331 ]\\
		&$\phi$ &0.992 & 0.944 & 0.402 & [ 0.368 ; 1.932 ]\\
		&$\beta$ &-0.826 & -0.803 & 0.347 & [ -1.582 ; -0.229 ]\\ \hline
	\end{tabular}
	\caption{Summary measures of the posterior densities of the parameters of  EPS and DPS models
		for the Galicia moss data in 1997.}\label{tab:aplic:galicia:infB:modelo_}
\end{table}

\begin{figure}[hbt!]
	\centering
	\subfigure{\includegraphics[scale=0.32]{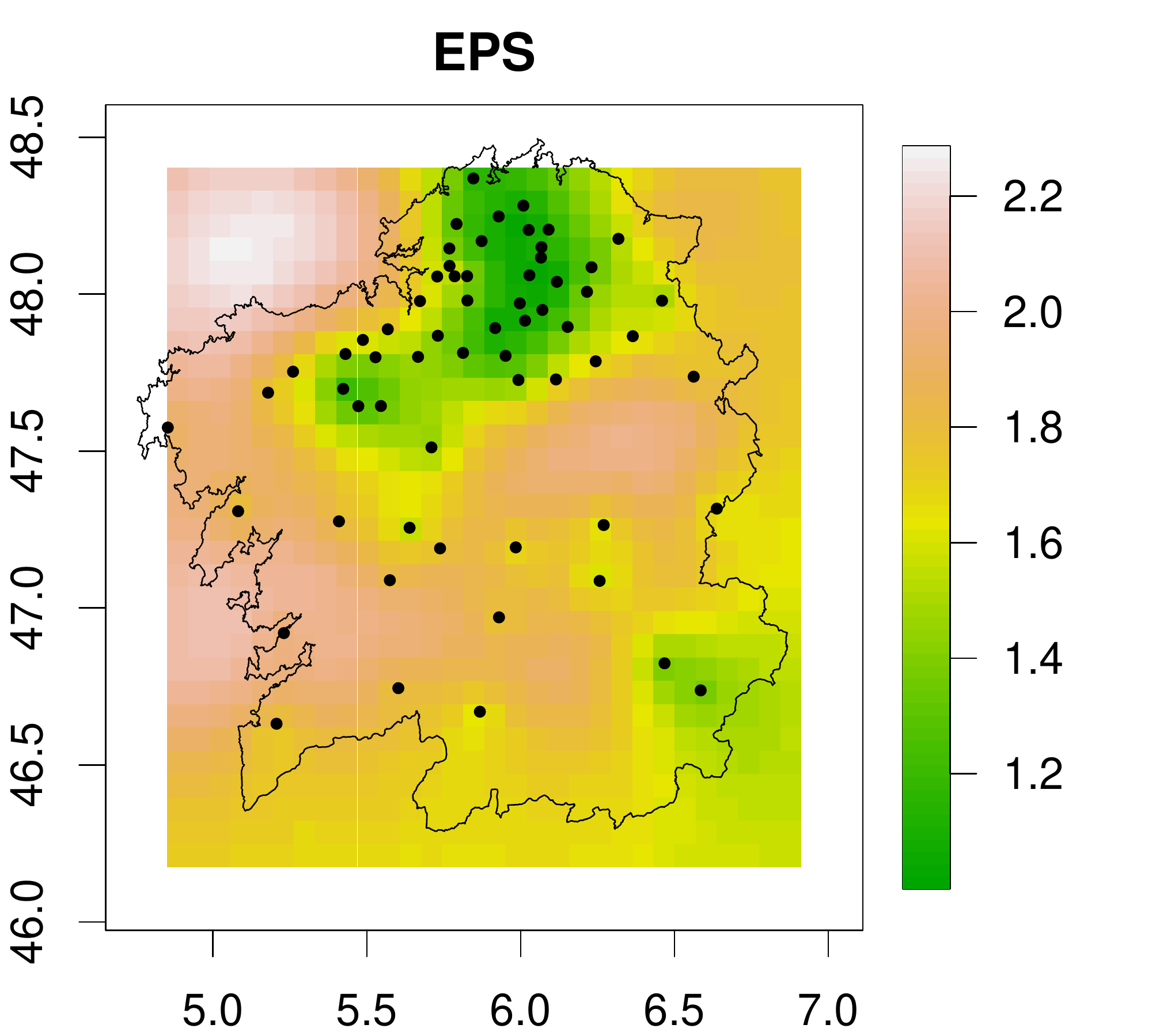}}
	\subfigure{\includegraphics[scale=0.32]{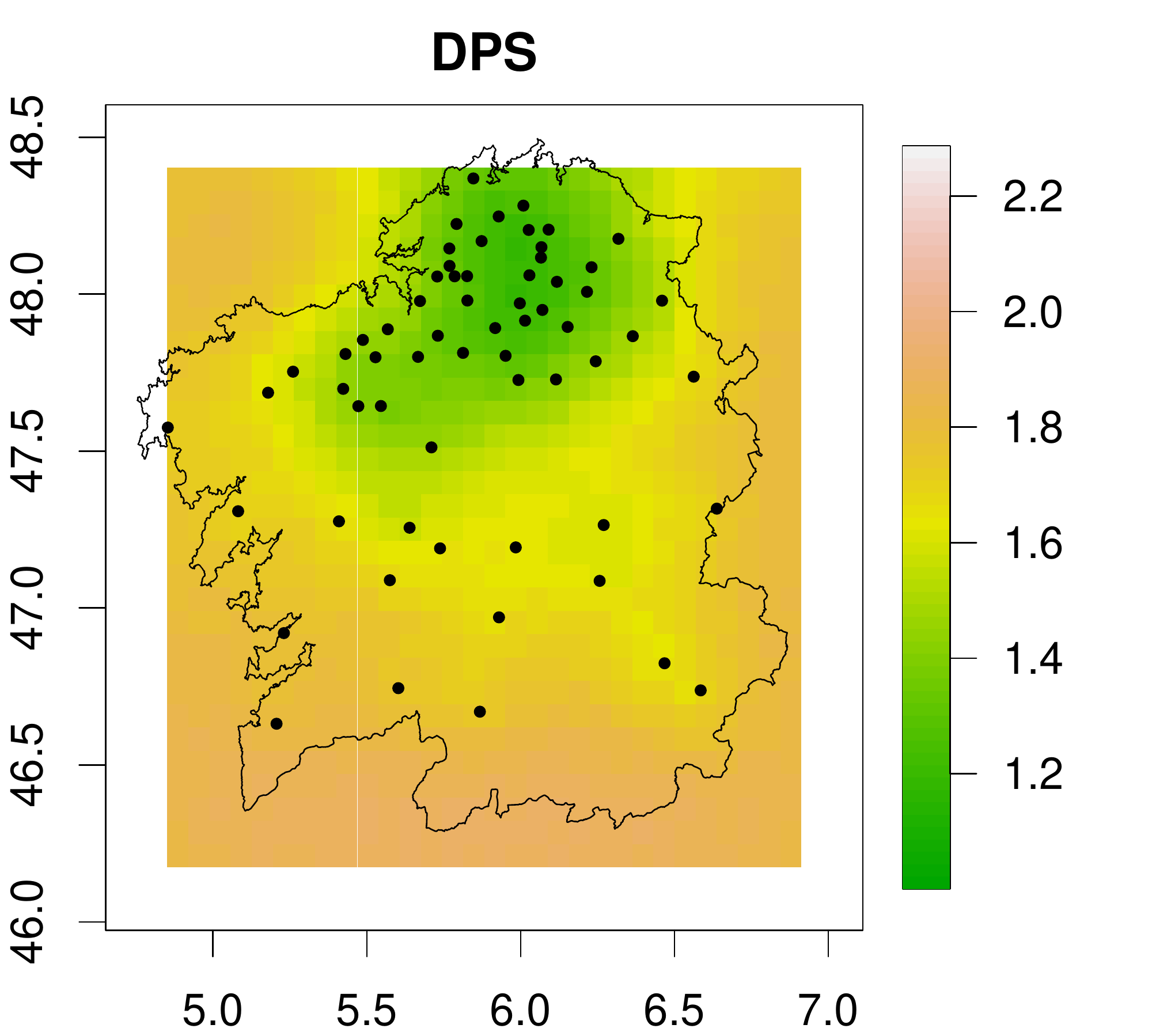}}	
	\caption{Predicted map of $Y$ of Galicia moss data in 1997 considering the EPS and DPS models.
		\label{fig:aplic:moss:pred97_loglead_infB}}
\end{figure}

Figure \ref{fig:aplic:moss:pred97_loglead_infB} shows the predicted maps in Galicia region over a regular grid of $30\times 30$.
The DPS model provided a shorter range of predicted values and lower prediction in regions without observations in comparison to the EPS model.

A cross validation was performed to compare the quality of the predictions of the EPS and DPS models.
Results are presented in Table \ref{tab:galicia:vc}.
The PPD measure corresponds to the sum of the log posterior predicted densities at the true value of the removed observation, summed up for all observations.
The EPS model provided better results in all measures, which evidences superior performance
of the exact approach in relation to the discretized approach.

\begin{table}[hbt!]
	\centering \renewcommand\arraystretch{1.2}
	\begin{tabular}{lccccc}
		\hline
		\multirow{2}{*}{Model}&\multirow{2}{*}{MAPE}&\multirow{2}{*}{PPD} & \multicolumn{3}{c}{CRCI} \\
		&&&90\%&95\%&99\%\\
		\hline
		EPS&0.388&-41.023&0.921&0.968&0.984\\
		DPS&0.415&-45.170&0.937&0.984&1.000\\
		\hline
	\end{tabular}
	\caption{Results of the cross validation exercise to the Galicia moss data.}\label{tab:galicia:vc}
\end{table}

\section{Discussion}\label{sec:discussion}

It is well known that inference can be incorrect if  sampling 
preferentiality is ignored in the analysis.
Most proposals to address preferentiality in geostatistics make use of model approximation.
This paper proposed a methodology to make exact inference for the geostatistical model
under preferential sampling without model approximations.

Results showed that the exact model provided good parameter  estimation and prediction in the
preferential and non preferential sampling context. In all simulated datasets, the estimation
values of the preferentiality parameter $\beta$ reflected the correct context of the data: values
different from zero for the former and values around zero for the latter. It has also
been shown that better results were obtained with the exact model in comparison against the
approximated model.

The proposed model was tested in two real datasets.
The first dataset is well known in the geostatistics literature as an example of non-preferential sampling. In this
situation, results showed that the exact model provided similar estimation and the posterior
distribution indicated that the model was able to identify the non-preferentiality of the data.
Also, similar prediction results were obtained with the proposed model, which evidences
the robustness of the exact approach even in the non-preferential context.
The second dataset is also well known in the geostatistics literature but as an example of preferential sampling.
Our results show that the models were capable to identify the preferentiality of the data but, more importantly,
the exact model outperformed the discretized model, confirming results obtained with simulations.

All exercises were based on the exponential correlation function. Other functions can be easily accomodated with the
appropriate modification of the corresponding step in the MCMC for sampling unknown hyperparameters of the correlation function.

Thus, there are strong reasons to believe that the proposed methodology can be used not only to correct the bias of the preferential sampling,
but also to verify if it exists in the data under analysis. Moreover, generalizations of the model
can be constructed to allow the analysis of non-Gaussian data.
This can be done by changing the distribution of the latent process and/or the response process.
Situations where outliers are present in data, for example, need distributions that are more
robust than the Gaussian and this feature can be of interest.

The assumption of gaussianity of the response can be easily relaxed to allow for other sampling data distributions.
This can be readily accommodated for a large number of distributions: scale mixtures of gaussian (including  t-Student), bernoulli and Poisson are just a few examples. Different augmentation schemes around the gaussian specification were designed for each one of them (see for respective details \cite{Carlin1992}, \cite{albert&chib1993}, \cite{Schnatter2006}).
All it takes is an additional, inexpensive draw associated with each observed response $Y$.

%
%

Finally, the methodology has substantial computational cost directly associated with handling high-dimensional multivariate normal distributions. Simulation of these distributions will be required and the computation becomes slow when the number of observed locations is large.
This is a well known problem in the spatial statistics literature and 
substantial amount of work was generated to provide adequate
solutions for these scenarios (see \cite{banerjee2008,kaufman2008,lindgren2011,datta2016}).
Improving the computational efficiency of our methodology is an important extension of this work.
\\~\\

\section*{Supplementary Material}
	Supplementary material for the article Exact Bayesian Geostatistics under Preferential Sampling. 
	This supplementary material provides proofs for some of the results and the remainder of the computational results of the simulated study and the application.

\noindent ACKNOWLEDGEMENTS

\noindent The authors acknowledge the partial financial support from CAPES-Brazil and CNPQ-Brazil.
The work is based on the doctoral thesis of the first author, developed under supervision of the
second author. Both authors thank the hospitality and the fruitful working environment of the Graduate
Program in Statistics at UFMG.
\vskip 3mm

{\bibliographystyle{jasa}
	\renewcommand{\refname}{\normalsize References}
	\setlength{\bibsep}{0.2pt}
\bibliography{sample.bib}}

\end{document}